\documentclass[twocolumn,preprintnumbers,amsmath,amssymb,prb,floatfix,superscriptaddress, nofootinbib]{revtex4-2}
\usepackage{graphicx}
\usepackage{dcolumn}
\usepackage{bm}
\usepackage{float}
\usepackage{color}
\usepackage[dvipsnames]{xcolor}
\usepackage{hyperref}
\usepackage{graphicx}
\usepackage{braket}
\usepackage{appendix}
\usepackage{chemmacros}
\usepackage{multirow}
\usepackage{natbib}
\usepackage{amsmath}
\usepackage{xcolor}
\hypersetup{colorlinks=true, citecolor=blue, urlcolor=blue, linkcolor=blue}
\usepackage{soul}

\newcommand{\be}{\begin{equation}}
\newcommand{\ee}{\end{equation}}

\newcommand{\bs}{\boldsymbol}
\usepackage{multirow}

\newcommand{\Vud}{V_{\uparrow \downarrow}}

\begin{document}   

\title{Topological Hall effect due to electron-skyrmion scattering}

\author{Arijit Mandal}{}\affiliation{Condensed Matter Theory and Computational Lab, Department of Physics, IIT Madras, Chennai-600036, India}
\affiliation{Center for Atomistic Modelling and Materials Design, IIT Madras, Chennai-600036, India}
\affiliation{Department of Physics \& Astronomy, University of Missouri, Columbia, MO 65211, USA}
\author{Hareram Swain}{}\affiliation{Condensed Matter Theory and Computational Lab, Department of Physics, IIT Madras, Chennai-600036, India}
\affiliation{Center for Atomistic Modelling and Materials Design, IIT Madras, Chennai-600036, India}
\author{B. R. K. Nanda}\altaffiliation[nandab@iitm.ac.in]{}\affiliation{Condensed Matter Theory and Computational Lab, Department of Physics, IIT Madras, Chennai-600036, India}
\affiliation{Center for Atomistic Modelling and Materials Design, IIT Madras, Chennai-600036, India}
\author{S. Satpathy}\altaffiliation[satpathys@missouri.edu]{}\affiliation{Condensed Matter Theory and Computational Lab, Department of Physics, IIT Madras, Chennai-600036, India}
\affiliation{Center for Atomistic Modelling and Materials Design, IIT Madras, Chennai-600036, India}
\affiliation{Department of Physics \& Astronomy, University of Missouri, Columbia, MO 65211, USA}    

\begin{abstract}

Electron scattering from chiral spin textures such as skyrmions is fundamental to the understanding of transport in more complex systems, 
including skyrmion crystals. 
Most of the previous studies have focused on the weak-coupling regime, where the exchange interaction is small compared with the electron energy. 
Real materials, however, often lie in the strong-coupling regime, which exhibits qualitatively different behavior. Using the Lippmann–Schwinger equation and Green's function formalism, valid for all coupling strengths, we uncover several new features in the scattering cross section, including Ramsauer–Townsend minima, pronounced intermediate-coupling resonances, and Landau-level resonances for skyrmions with larger winding numbers. 
These features strongly influence the topological and spin Hall conductivities, which depend sensitively on the incident electron energy. 
Our work provides important insights into the  Hall transport in collective chiral spin textures such as the skyrmion crystal.
\end{abstract}   
\maketitle

\section {Introduction}

The non-collinear chiral magnetic textures such as the skyrmions constitute an important class of topologically nontrivial states in condensed matter physics \cite{Bogdanov1, Bogdanov2, skyrmion1, skyrmion2, skyrmion3, skyrmion4}. 
This has  opened up a new research area for fundamental science as well as for device applications 
due to their nanoscale size, particle-like mobility, metastability, and unusual electrodynamics. 
The emergent magnetic field in these structures due to the non-collinear spin texture  drives the itinerant electrons in the transverse direction, resulting in novel electron transport such as the topological Hall\cite{Nagaosa_the, Mandal}, spin Hall\cite{Goebel_spin}, and orbital Hall\cite{Goebel_orbital} effects.  

There have been numerous theoretical and experimental studies of transport in skyrmions and other noncollinear spin textures, particularly in collective systems such as the skyrmion crystals \cite{nagaosa-review, tokura-review, thc-review, Beyond, the1, the2, the3, the4, Nagaosa_the, Mandal, Goebel_spin, Goebel_orbital}. Electron scattering from isolated spin textures, such as a single skyrmion, is a fundamental building block for understanding transport in these collective phases. Yet, only a limited number of studies have examined scattering from an individual skyrmion, and most have focused on the weak scattering potential regime\cite{Denisov-prl, Denisov-sci-rep, Denisov-prb-1, Denisov-prb-2, Denisov-jpcc, Rashba-Ramsauer, Blugel-prl, Hareram}. 
The scattering potential is governed by the exchange interaction $J$ between the itinerant electron spin and the local skyrmion spins, so that in these works $J$ is assumed to be sufficiently small as compared to the electron energy for perturbative Born approximations to remain valid. Despite this limitation, these studies have established several important concepts governing electron scattering from noncollinear spin textures. 

However, in most real materials, including many skyrmion-hosting systems, the exchange interaction $J$ is comparable or even much higher as compared to the typical electron energy, and it is in fact often approximated as $J \rightarrow \infty$\cite{Nagaosa_the, Mandal, Goebel_spin, Goebel_orbital}. 
To address this issue, in a recent work\cite{Hareram}, we studied the
dynamics of the electron-skyrmion scattering  from the time-evolution of a propagating Gaussian wave packet by solving the time-dependent Schr\"odinger equation.
The method reveals the intricate dynamics of the scattering process as the electron moves through the skyrmion region; however, due to the computational limitations that allows simulations for limited propagation times, a different approach such as the one presented here is needed for an accurate determination of  the long-distance behavior such as the scattering cross section and the Hall conductivities.

\begin{figure}[hbt!]
    \centering
    \includegraphics[width=0.9\linewidth]{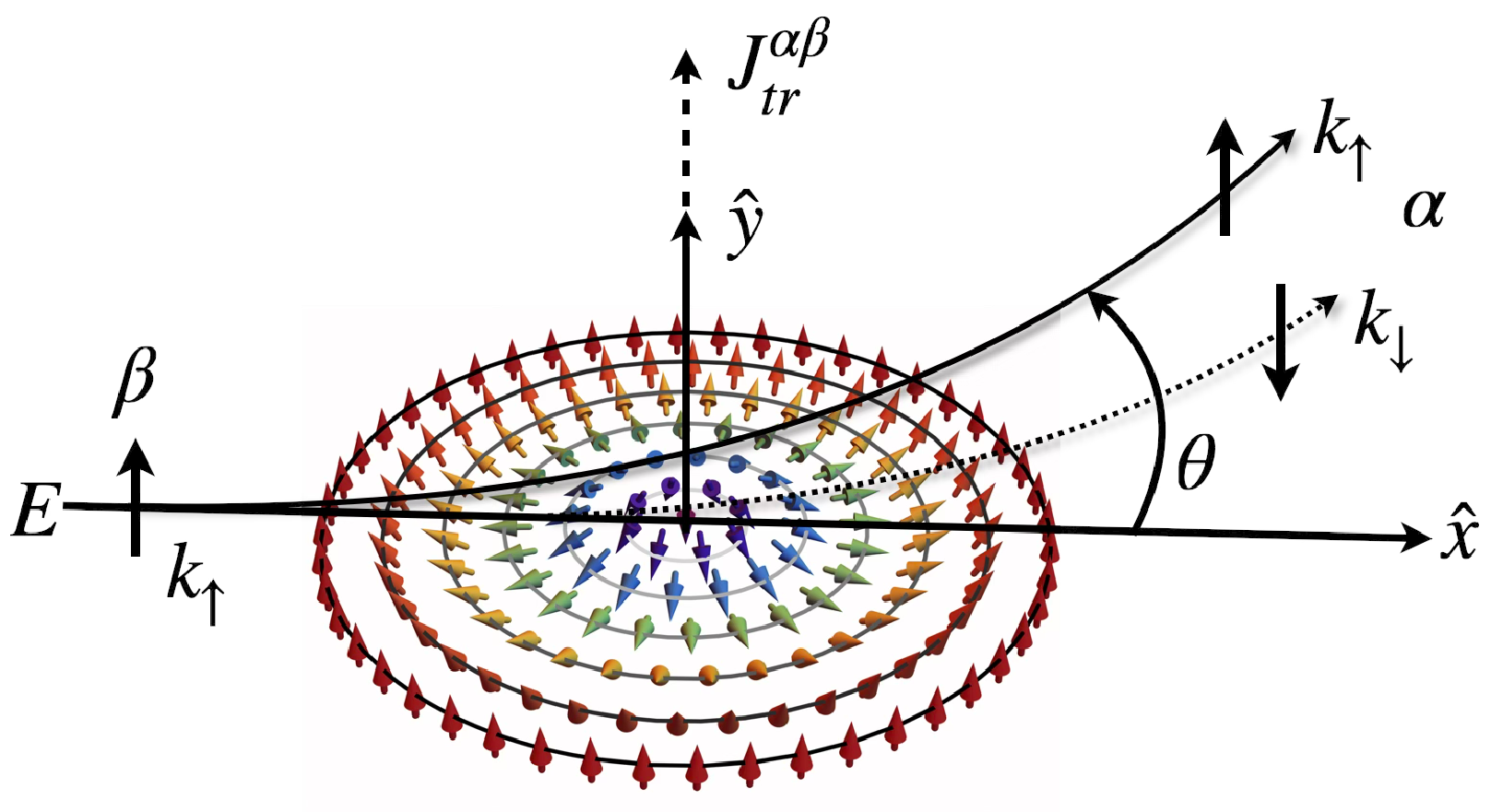}
    \caption{ 
    Schematic diagram of scattering of a spin up electron. 
  The scattered wave consists of both a spin-preserved component and a spin-flip component with spin-dependent momentum $k_\sigma$. 
  The net transverse current is denoted by $J_{tr}^{\alpha\beta}$, where $\beta$ ($\alpha$) is the spin of the incident (scattered) electron.
  For energy $E < J$, even though there is no spin down incident or scattered beam, the spin down state can  nevertheless  exist in the skyrmion region, produced via the spin-flip potential, and can thereby influence the scattering process.
    }
    \label{Fig1}
\end{figure}

In this paper, we study the electron-skyrmion scattering process from the solution of the Lippmann-Schwinger equation and by evaluating the full Green's function (GF).  
Spin-resolved transverse Hall currents and the scattering cross sections are studied, from which we compute the topological charge Hall and spin Hall conductivities.
Our formulation is valid for all strengths of the scattering potential, characterized by the exchange interaction $J$ between the electron and the skyrmion spins. 
While our results agree with the scattering theories \cite{Denisov-prl, Blugel-prl} in the weak exchange limit, they reveal new scattering features for higher values of $J$ such as the low energy divergence, peaked structures and the Ramsauer-Townsend dips
for intermediate $J$, as well as resonance peaks originating from the Landau levels for skyrmions with higher winding numbers.
Another noteworthy result is that by tuning the energy of the incident electrons, the transverse charge current can be suppressed resulting in a pure spin Hall current,
which may be of potential value for generating spin currents in spintronics devices.

\section {Scattering Theory from the single-center spin texture}
\label{sec:theory}

We consider scattering from a two-dimensional (2D) spin texture defined 
by the orientation of the local spins $\bs {\hat{S}} (r, \theta)$, where 
$(r, \theta)$ are the polar coordinates on the 2D plane. 
We assume that the local spins as $r \to \infty$ are oriented along $+z$ axis, i.e.,$ \lim_{r \to \infty}  \bs {\hat S} ( r,\theta) = \bs{\hat z} $, as appropriate for the skyrmions. 
The Hamiltonian  for the electron moving in the spin texture
is written as
\begin{equation} 
{\cal {H}} = \frac{p^2}{2m_e} + \frac{J}{2}   ( 1- \bs { \hat S } \cdot \bs { \sigma} ),
\label{Eq1}
\end{equation}
where $J$ is the Hund's energy, $\hat {\bs S}$ is the local spin orientation, and $\bs \sigma = (\sigma_x, \sigma_y, \sigma_z)$ are the Pauli matrices describing the electron spin. Spin interaction energy has been defined to be zero if the electron spin is aligned parallel to the local spin, and $J$ if it is anti-parallel. 
It is convenient to rewrite the Hamiltonian as ${\cal {H}} = {\cal {H}}_0 + V$, where 
\begin{equation} 
{\cal {H}}_0 =  \frac{p^2}{2m_e} + \frac{J}{2}   ( 1 - \sigma_z )
= \begin{pmatrix}
        p^2/2m_e & 0 \\
        0 & p^2/2m_e + J
         \end{pmatrix},
         \label{Eq:H0}
\end{equation}
and the scattering potential is 
\begin{equation}
    V(\bs r) \equiv \frac{J}{2}   ( \sigma_z - \bs { \hat S ( r) } \cdot \bs \sigma ) = \begin{pmatrix}
        V_{\uparrow \uparrow} & V_{\uparrow \downarrow} \\
        V_{\downarrow \uparrow} & V_{\downarrow \downarrow}
         \end{pmatrix}.
         \label{Eq:vsc}
\end{equation}
Note that with this partition, $V (\bs r)$ is non-zero only in the scattering region $r < \lambda$,  for both spin up or down incoming particles, since  $  \hat {\bs S} (\bs r) \parallel \hat {\bs z} $ outside this region. Thus the problem can be formulated in terms of the standard scattering theory.
We consider the scattering of the incident plane wave 
\begin{equation}
   | \phi_{k\sigma} \rangle = e^{i\bs k_\sigma \cdot \bs r} |\sigma \rangle,
\end{equation}
where we take the spin state $|\sigma \rangle$  to be either parallel or anti-parallel to the background spin configuration. 
 For the incident energy $E$, the momentum $k_\sigma$ is given by the expressions $E = \hbar^2 k_\uparrow^2 /2m$ and $E = \hbar^2 k_\downarrow^2 /2m + J$. 
 
 The scattered wave function  is obtained from the Lippmann-Schwinger equation, written in terms of the $T$ matrix as
\begin{equation}
 |\psi_{k\sigma} \rangle = |\phi_{k\sigma} \rangle + G_0 T |\phi_{k\sigma} \rangle,
 \label{Eq:LS}
 \end{equation}
where $T (E)  = V (1-G_0(E) V)^{-1}$, $G_0 (E) = (E-{\cal {H}}_0+i \eta)^{-1} $ is the retarded GF, corresponding to the outgoing boundary condition,  and $E$ is the energy of the electron both before and after  scattering  (elastic scattering). The spin-momentum indices in $\phi_{k \sigma}$ label the incoming electron. 
The scattered wave  function $|\psi_{k\sigma}\rangle$ in Eq. (\ref{Eq:LS}) will contain a spin piece with the original spin-momentum $(k_\sigma, \sigma)$ and a spin-flip piece with spin-momentum $(k_{-\sigma}, -\sigma)$  as illustrated in Fig. \ref{Fig1}. 

 We are interested in the long-distance behavior of $|\psi_{k\sigma} \rangle$, far away from the scattering center,
 which follows from the long distance behavior of $G_0$.
 For free particles, $G_0$ is spin diagonal, and  for the 2D case, it is given\cite{Sadhan} in terms of the Hankel functions. The result is
 \begin{eqnarray}
    G_0^{\alpha \beta}(\boldsymbol{r},\boldsymbol{r'},E) 
    =
    - \frac{2m_e}{\hbar^2}\frac{i}{4} H_0^{(1)}(k_\alpha |\boldsymbol{r}-\boldsymbol{r'}|)\  \delta_{\alpha \beta},
    \label{Eq6}
\end{eqnarray}
where $\alpha, \beta$ are the spin indices and $G_0^{\alpha \beta}(\boldsymbol{r},\boldsymbol{r'},E) 
    \equiv \langle \bs{r}\alpha| G_0(E)| \bs{r'}\beta \rangle$\cite{Note2}.
Since the scattering potential is limited to a finite region near the origin ($r  < \lambda)$,
the asymptotic limit of Eq. (\ref{Eq6}) determines the scattering cross section. The limit is
\begin{equation}
     \lim_{r \to \infty} G_0^{\alpha\alpha} (\bs{r},\bs{r'}, E) = -\frac{2m_e}{\hbar^2}\frac{e^{i\pi/4}}{4} \sqrt{\frac{2}{\pi k_\alpha r}} e^{ i k_\alpha r} e^{- i \bs{k'}_\alpha \cdot \bs{r'}},
     \label{Eq:G0-limit}
\end{equation}
where $\bs{k'}_\alpha \equiv k_\alpha \bs {\hat r} $\cite{Morse-Feshbach}.
Substituting Eq. (\ref{Eq:G0-limit}) into the Lippmann-Schwinger equation (\ref{Eq:LS}) and considering scattering of  the spin eigenstates $|\sigma \rangle$ of $S_z$, 
the asymptotic result for the scattered wave function is
\begin{equation}
    \psi_{k\sigma} (\bs r) \xrightarrow [r \to \infty] {} 
    e^{i\bs k_\sigma \cdot \bs r} |\sigma \rangle + \sum_{\sigma'} \frac{e^{ i k_{\sigma'} r}}{ \sqrt { r}}
    f_{\sigma' \sigma}^{k_\sigma,k_{\sigma'}} (\hat r) \ |\sigma' \rangle.
        \label{Eq92}
\end{equation}
The exponential factor in the second term indicates that the scattered waves of different spins travel with different speeds.
The scattering amplitude  is given by the $T$-matrix
\begin{equation}
     f_{\alpha\beta} (\hat r) = -  \frac{m_e}{\hbar^2} \sqrt { \frac{i}{2\pi k_\alpha}  } \langle \bs k_\alpha^\prime \alpha| T |
     \bs {k}_\beta \beta \rangle,
     \label{Eq:f}
\end{equation}
where we have suppressed the superscripts  in 
$f_{\sigma' \sigma}^{k_\sigma,k_{\sigma'}}$ and
\begin{equation}
   \langle \bs k_\alpha^\prime \alpha| T |
     \bs {k}_\beta \beta \rangle = 
     \int d^2 r' d^2 r''\ 
     e^{- i \bs {k'}_\alpha \cdot \bs {r'} }   e^{ i \bs {k}_\beta \cdot \bs {r''} } T_{\alpha \beta} ( \bs r', \bs r'').
     \label{Eq:Tkk}
\end{equation}
The $T$-matrix elements are obtained by first inverting the GF matrix in real space:
\be
T_{\alpha \beta} ( \bs r', \bs r'') = \langle \bs r' \alpha| V (1-G_0 V)^{-1} | \bs r'' \beta \rangle, 
\label{Eq12}
\ee
and then by taking the double Fourier transform as indicated in Eq. (\ref{Eq:Tkk}).
The  total scattering cross section is, as usual,
$ \sigma_{\alpha \beta} = \int_0^{2\pi} (  d\sigma/d\theta)_{\alpha \beta} \ d\theta  $, where  $(d\sigma/d\theta)_{\alpha \beta} \equiv |f_{\alpha \beta} (\theta)|^2$ is the differential scattering cross section.

In the Born approximation, the $T$ matrix appearing in Eq. (\ref{Eq:LS}) is $T = V$, while in the second Born approximation, the next term is kept in the series expansion, so that $ T = V + V G_0 V$.
Since the electron must scatter from at least a triad of skyrmion spins $\bs S$ for
topological effects, 
 topological differences in the scattering cross section will show up in terms $O(V^3)$ and higher\cite{Denisov-prl}. Thus, at least the second Born approximation must be retained to describe the topological effects. In our calculations, we have used the full GF, so that terms to all orders in $V$ are included, and our results are valid for all scattering parameters without any approximation. The $T$ matrix  is computed in real space, following Eq. (\ref{Eq12}), by direct matrix inversion on a two-dimensional grid  of $30 \times 30$ points in the polar coordinates covering the skyrmion region.  

Note that, in an important practical situation ($E < J$), which can happen in typical solids,  momentum of the spin down electron is imaginary. Denoting it by $i k_\downarrow$, where $k_\downarrow$ is real, the corresponding propagator, Eq. (\ref{Eq6}), has an exponentially-decaying behavior, coming from the Hankel function of imaginary argument. In this case, analogous to Eq. (\ref{Eq:G0-limit}), we have 
\begin{equation}
     \lim_{r \to \infty} G_0^{\downarrow\downarrow} (\bs{r},\bs{r'}, E) = -\frac{m_e}{2\hbar^2}  \sqrt{\frac{2}{\pi k_\downarrow r}} e^{ - k_\downarrow r} e^{ \bs{k'}_\downarrow \cdot \bs{r'}},
     \label{Eq:G0-limit2}
\end{equation}
where $ {\bs k'}_\downarrow =  k_\downarrow \bs{ \hat r}$, which leads to an exponentially-decaying evanescent wave away from the scattering center, $r \gg r'$. Though  the spin down electrons cannot exist outside the skyrmion if $E < J$, they nevertheless can exist in the skyrmion region and influence the scattering process. A spin up electron can produce a spin down electron via the spin-flip potential term $V_{\downarrow \uparrow}$ within the skyrmion region, which however needs to be back converted to spin up before the electron can escape.

\begin{figure}[hbt!]
    \centering
    \includegraphics[width=1.0\linewidth]{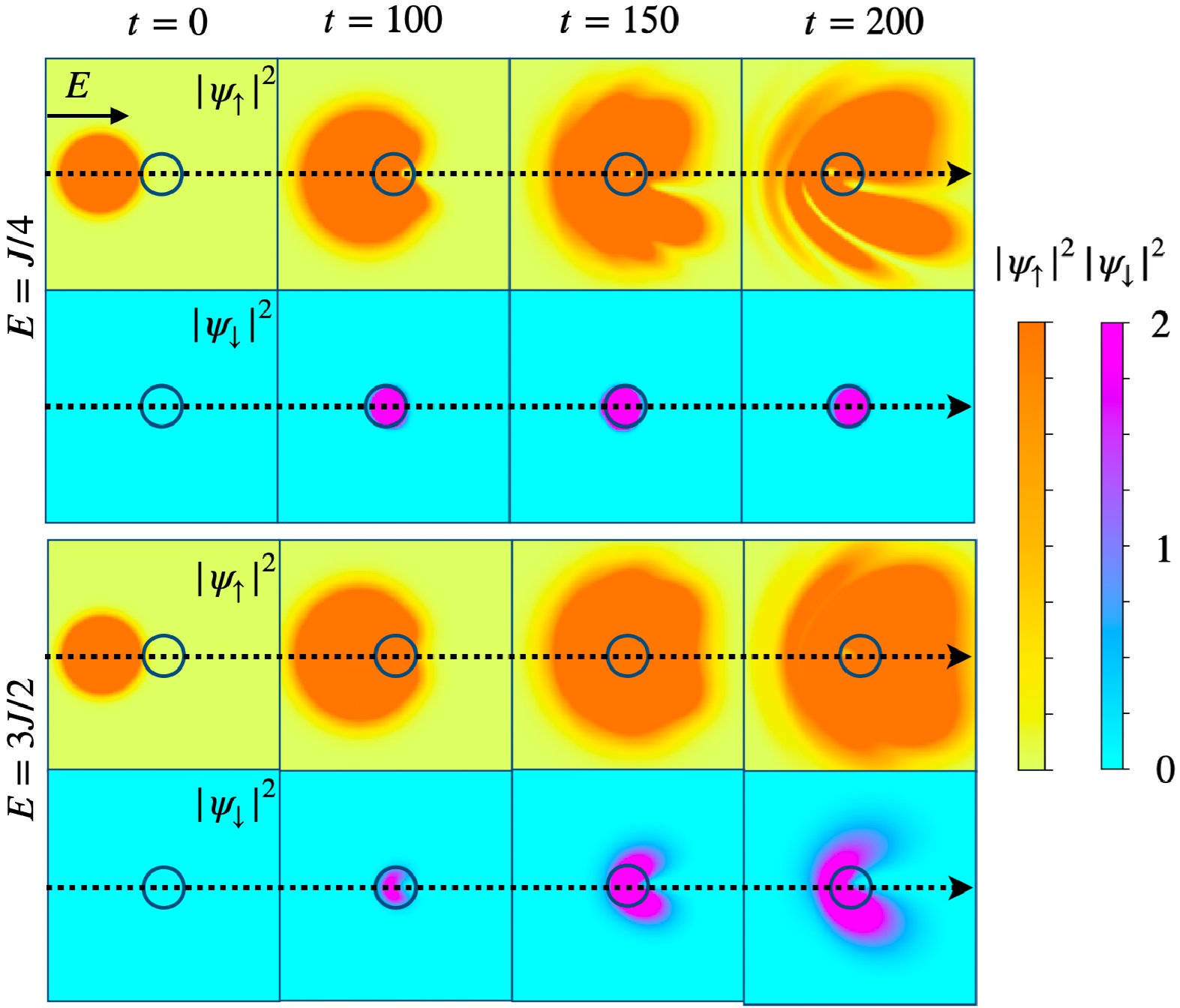}
    \caption{Scattering dynamics of a spin-up Gaussian wavepacket from numerical solution of the time-dependent Schr\"odinger equation following the methods of Ref. \cite{Hareram}.
    The {\it upper panel} indicates the trapped spin-down state for $E < J$,
  dynamically generated inside the skyrmion, indicated by the black circle, during the scattering process.
   In contrast, the {\it lower panel}, where $E > J$, shows a small propagating spin down component after scattering, which will result in a spin-down current at $r \rightarrow \infty$. All figures are 
   time snapshots of probability distribution $|\psi|^2$ in both spin up and down channels as indicated. The color coding, which is different for the two spin channels, indicates the magnitude of  $|\psi|^2$. The color intensity map and the time steps are both in arbitrary units, but in all cases the normalization of the wave function is enforced. The figure also illustrates the asymmetric nature of the scattering (left of the scatterer vs. right) which leads to the charge Hall and the spin Hall effects.
   }
    \label{Fig2}
\end{figure}

As an aside, it is interesting to see this point from a direct numerical simulation of the
time-dependent spinful Schr\"odinger equation for the scattering of a Gaussian wave packet from the skyrmion potential. The details of the method are discussed elsewhere\cite{Hareram}.
Fig. \ref{Fig2} shows the time snapshots during the scattering process for a spin up incident electron with two different energies. For the {\it lower panel}, the electron energy is $E > J$, which means that both spin up and down electrons can exist as propagating states. After the scattering, the electron propagates mostly as a spin up state, but there is a small propagating spin down component. In contrast, when $E < J$ ({\it upper panel}), only the spin up propagating state is possible. The spin-down state generated inside the skyrmion during the scattering process continues to stay as a trapped state for a very long time, since it can only escape by a back conversion to a spin up state. In fact, for the entire duration of the simulation, the spin down state remains trapped within the skyrmion (magenta color in the second panel from top in Fig. \ref{Fig2}).

\begin{figure}[tbh!]
    \centering
   \includegraphics[width=0.9\linewidth]{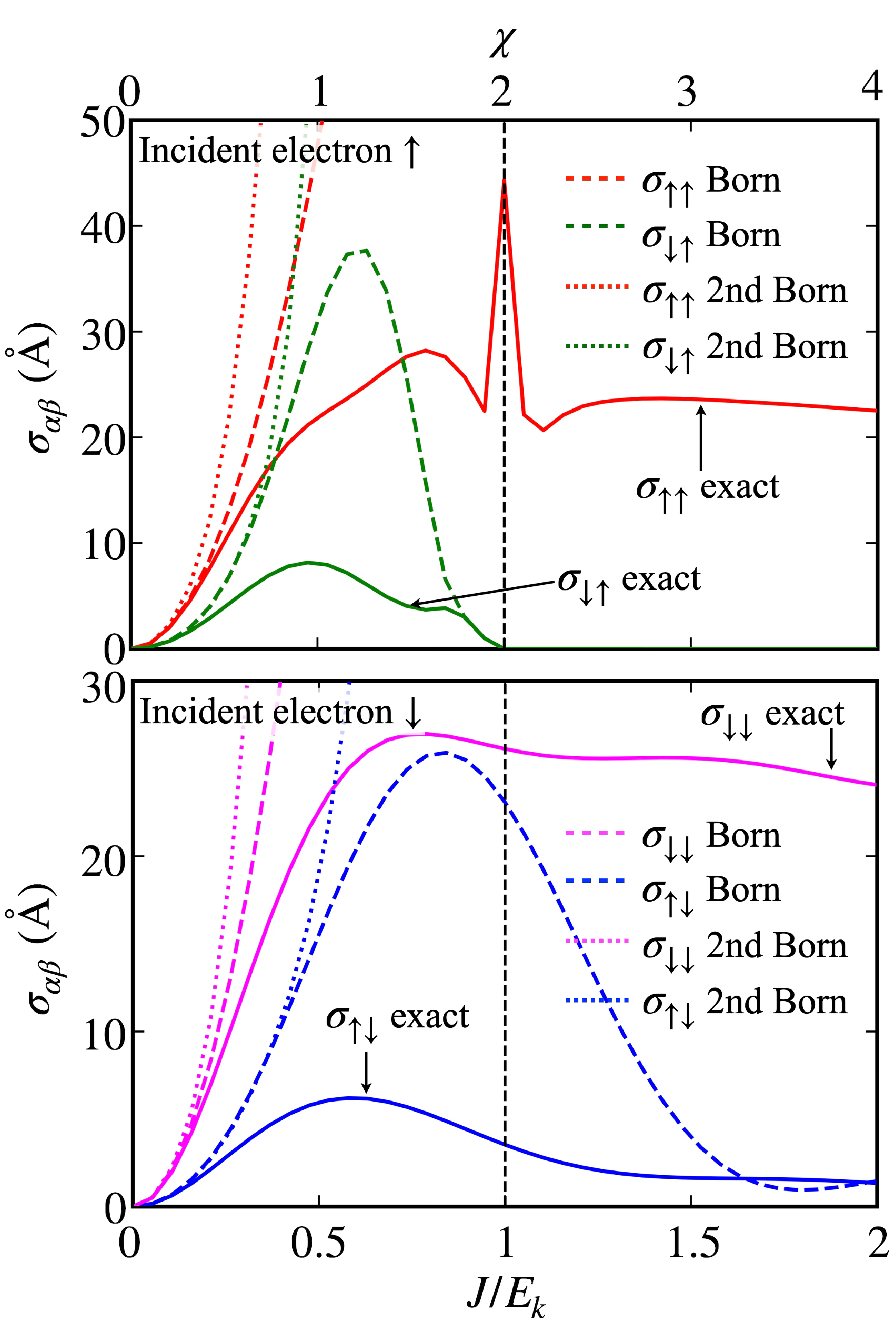}
    \caption{
Total scattering cross section $\sigma_{\alpha \beta}$ for a skyrmion with winding number $m = 1$ as a function of the exchange energy $J$ (the corresponding adiabatic parameter $\chi$ is shown on the top). Born approximation and exact results are shown for both  spin up ({\it {top}}) and  spin down ({\it {bottom}})  incoming electron.  The second Born approximation contains the  three-spin scattering processes to the lowest order and hence the effect of chirality. However, the Born approximation results deviate quickly from the exact results as $J$ is increased, necessitating a full solution for typical solids.
In the figure, the kinetic energy of the incident electron $ E_k = 4$ eV and the skyrmion radius $\lambda = 10$ \AA.}
    \label{Fig3}
\end{figure}

{\it Adiabatic parameter}.  It is convenient to define an adaibatic parameter $\chi$ for the scattering process. Adiabaticity implies that the electron has adequate time to change its spin state to that of the local spin. From the time-energy uncertainty principle, this time is $\tau = \hbar/J$, while the time taken to pass through the atom containing the local spin is $T = 2a/v$, where $a \approx 2$ \AA\ is the typical interatomic distance, and $v$ is the electron velocity. Thus the adiabatic condition means that the adiabatic parameter 
$\chi \equiv T/\tau \gg 1$, or
\begin{equation}
    \chi  \equiv T/\tau = \frac{Ja}{2 \sqrt {E_k}} \gg 1,
\end{equation}
where $a$ is in \AA, $J$ and  $E_k$ (the  kinetic energy of the incident electron) are in units of eV, and we have taken $ \hbar^2/ (2m_e) \approx 4$ eV.\AA$^2$. For typical solids, $J \sim 2$ eV, $E_k \sim 1$ eV, $a \sim 2$ \AA, so that $\chi \sim 2$, which makes the electron scattering  more or less adiabatic. In view of this, the approximation $J \rightarrow \infty$ is often adopted to describe many spin phenomena in solids without making any serious error.

For the electron-skyrmion scattering, often the small $J$ limit has been treated in the literature, and the Born approximation has been used to extract the effects of chirality\cite{Denisov-prl, Denisov-sci-rep, Blugel-prl}. However, for realistic parameters, many qualitative aspects of the scattering process are not described in the small $J$ approximation. 
As seen from Fig. \ref{Fig3}, the Born approximation agrees with the exact results in the small $J$ limit as expected,  but fails to describe the scattering process for realistic parameters in the solid, where $J$ is large and the scattering is adiabatic ($\chi \gg 1$).
Our formulation, being exact, is valid for the entire range of  parameters.

\section{ Scattering from the skyrmion spin texture: Numerical Results}

\begin{figure}[tbh!]
    \centering
   \includegraphics[width=0.7\linewidth]{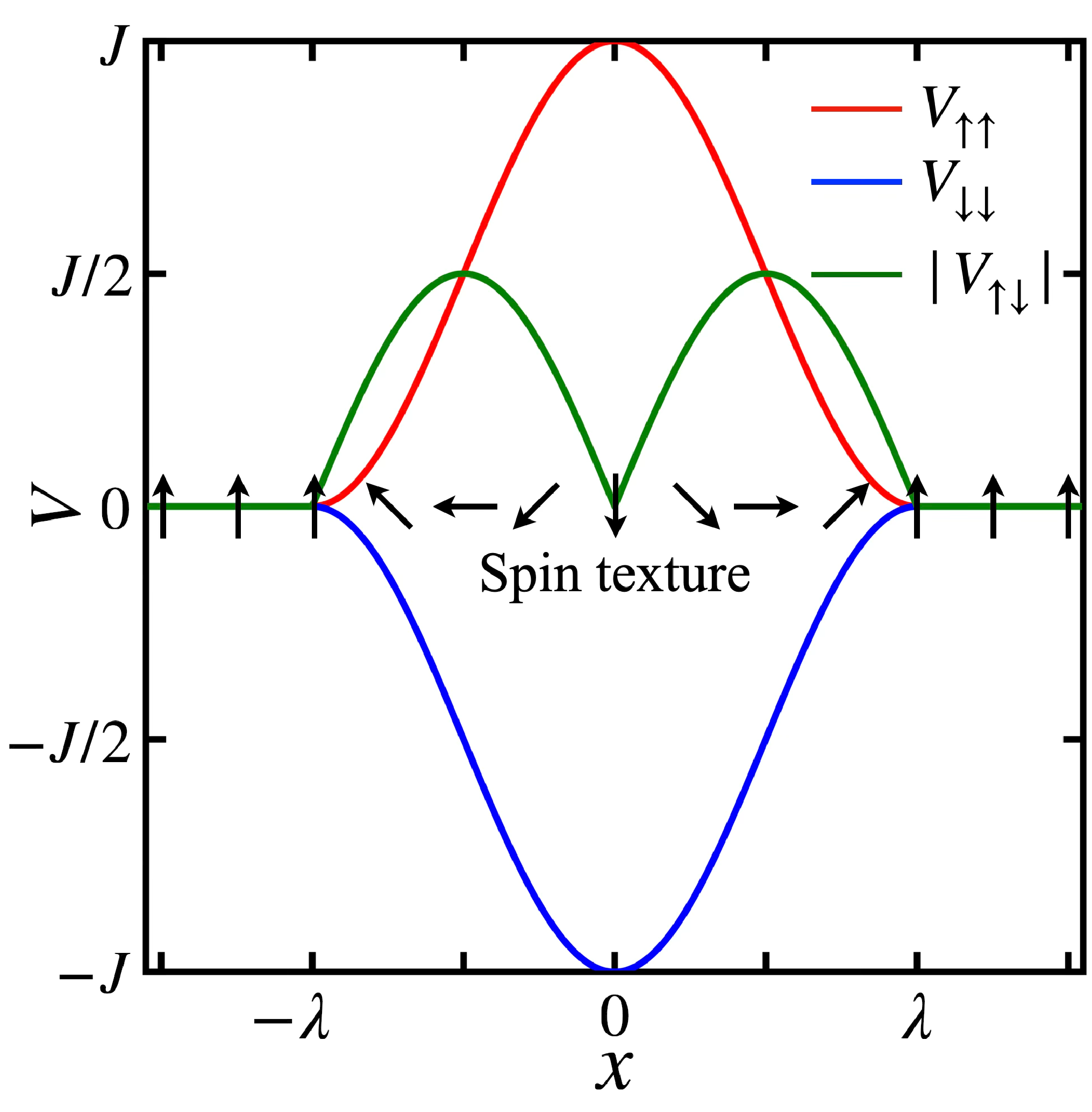}
    \caption{
Scattering potential  $V (r)$ for the skyrmion corresponding to Eq. (\ref{Eq:potential}) plotted along the $x$ axis. Here, $\lambda$   is the skyrmion radius,  and $J$ is the exchange parameter. The spin up (down) electron experiences a spin-preserving repulsive (attractive) potential together with a complex spin-flip scattering potential, whose magnitude $|\Vud|$ has been shown. There is no scattering potential outside $ |r| > \lambda$, and the arrows on the central line indicate the skyrmion spins.
}
    \label{Fig4}
\end{figure}

\begin{figure*}[tbh!]
    \centering
   \includegraphics[width=1.0\linewidth]{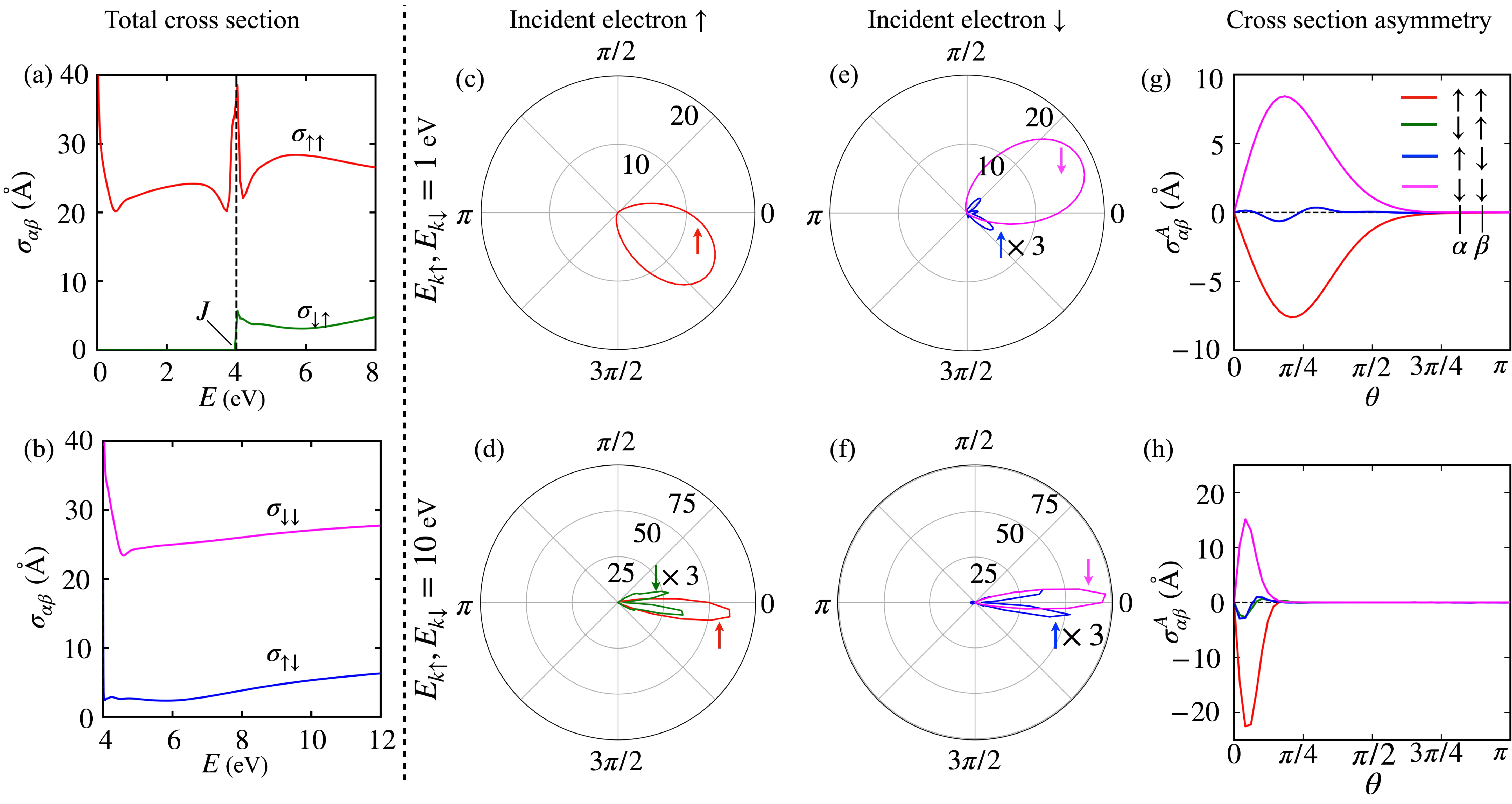}
    \caption{
   Typical features of the scattering cross section. (a) Total scattering cross section $\sigma_{\uparrow \uparrow / \downarrow \uparrow}$ for incident spin-up electron. (b) Same for the spin-down electron, denoted by $\sigma_{\uparrow \downarrow / \downarrow \downarrow}$. 
(c) Angular dependence of the differential scattering cross section for incident spin-up electrons with energy $E = E_{k\uparrow}= 1$ eV.  
Since    $E < J$, there is no spin-down scattering component, as it cannot exist outside the skyrmion region. (d) The same for $E_{k \uparrow} > J$, but here scattering can occur to both spin channels. 
(e) and (f) Differential scattering cross section for spin-down incident electrons with two different energies. In this case, scattering is allowed to both spin channels for all energies.  
The spin-flip scattering cross-sections have been multiplied by a factor of three for (c) - (f). Note from (c)-(f) that the incident spin-up electron scatters predominantly to the lower half plane ($\theta < 0$), while the spin-down electron scatters to the upper half plane ($\theta > 0$), caused by the chiral spin structure of the skyrmion. 
(g) and (h)
 Scattering asymmetry $\displaystyle {\sigma^{\rm A}_{\alpha \beta} (\theta)  \equiv (d\sigma_{\alpha \beta}/ d \theta) (\theta) - (d\sigma_{\alpha \beta}/ d \theta) (-\theta)}$ for the two different incident energies. 
 In all figures, parameters are: $ J = 4 $ eV, the skyrmion radius $\lambda = 10$ \AA, and winding number $m =1$. All scattering cross sections are color coded (as indicated in (a) and (b)), and $\sigma_{\alpha \beta}$, here and throughout the paper, indicates incident spin $\beta$ scattered into the  spin $\alpha$  channel.
}
    \label{Fig5}
\end{figure*}

\subsection { skyrmion spin texture}
For the numerical calculations, we use the skyrmion spin texture, commonly used in the literature, viz.,  
\begin{eqnarray}\label{Eq:skyrmion theta phi defn}
    \bar \theta(r) &=& 
    \begin{cases}
    \pi(1-r/\lambda) & \text{for}~ r \leq \lambda \\
    0 & \text{for}~ r > \lambda,
    \end{cases}\nonumber \\
    \bar \phi(\theta) &=& m\theta + \gamma,
    \label{texture-skx}
\end{eqnarray}
where $(\bar \theta, \bar \phi)$ define the orientation, in spherical coordinates, of the skyrmion spin located at  the polar coordinates $(r, \theta)$ on the 2D plane,
$m$ is the winding number of the skyrmion, and $\gamma$ is the helicity.
Thus, the local magnetization vector is $\boldsymbol{S}(\boldsymbol{r}) = 
\hat{x}\ \sin{\bar \theta}\cos{\bar \phi}  + \hat{y} \ \sin{\bar \theta}\sin{\bar \phi}  + \hat{z} \ \cos{\bar \theta} $ in cartesian coordinates, $\hat z$ being normal to the plane. The scattering potential in Eq. (\ref{Eq:vsc}) becomes
\begin{align}
 V(r,\theta)  =  
-\frac{J}{2} \begin{pmatrix}
     \cos{\bar \theta }-1 & \sin \bar  \theta \,e^{-i \bar  \phi}\\
     \sin \bar  \theta \,e^{i \bar  \phi}  & -\cos \bar  \theta + 1
   \end{pmatrix}.
   \label{Eq:potential} 
\end{align}
Since the helicity $\gamma$ appears as a constant phase factor in the off-diagonal elements of $V (r, \theta)$, it is easy to show that
it also appears as only a multiplicative overall phase factor in the $T$ matrix, Eq. (\ref{Eq12}), so that $\gamma$ does not affect the scattering cross section $d\sigma/d\theta \propto |T_{\alpha\beta}|^2$. We take $\gamma = 0$ in the numerical calculations.
Furthermore, since the scattering is characterized by a single parameter $J/E_k$ as seen from the Hamiltonian, Eq. (\ref{Eq1}), we have fixed the value $J = 4$ eV and varied the energy of the incident energy. We take the skyrmion radius   $\lambda= 10$ \AA\ and the winding number $m = 1$, unless otherwise stated.

\subsection{General features of the scattering cross section}
\label{IIA}
Fig. \ref{Fig5} summarizes the computed scattering cross sections for typical parameters. 
The spin-differentiated total scattering cross section $\sigma_{\alpha \beta}$, where $\beta$ ($\alpha$) denotes the incident (scattered) electron spin are shown in Fig. \ref{Fig5} (a) and (b). There are several features to notice: 

(1) The scattering cross section in the spin-preserved channel is much stronger than the spin-flip channel. 

(2) Only $\sigma_{\uparrow \uparrow}$ is non-zero for $E < J$ because a propagating spin down state (incident or scattered) exists only if $E > J$.

(3) For small kinetic energy ($E_{k\uparrow} = E$, $E_{k\downarrow} = E - J$), both $\sigma_{\uparrow \uparrow}$ and 
$\sigma_{\downarrow \downarrow}$ show similar divergence.

(4) For the energy shown in  Fig. \ref{Fig5}, the magnitude of $\sigma_{\uparrow \uparrow}$ and 
$\sigma_{\downarrow \downarrow}$ are comparable to the classical scattering cross section, which is for the 2D case, $\sigma = 2 \lambda = 20$ \AA. For  $ E \gg J$, the cross sections fall to zero (not shown in the figure).

(5) As seen from Fig. \ref{Fig5} (c)-(f), the incident spin-up electron scatters predominantly to the lower half plane ($\theta < 0$), while the spin-down electron scatters predominantly to the upper half plane ($\theta > 0$), caused by the chiral spin structure of the skyrmion. This is reversed if the winding number is reversed in sign.  

For $\sigma_{\uparrow \uparrow}$,
there are three noticeable features: 
(i) Divergence of  at $E_k =0$, 
(ii) A peak-like structure at $E_k \approx J$, and 
(iii) A dip in the scattering cross section on either side of the peak. 
For the $\sigma_{\downarrow\downarrow}$ scattering, the main noticeable feature is the divergence at $E = J$, which is similar to the $\sigma_{\uparrow\uparrow}$ channel for $E =0$ as already mentioned.

The main features of the $\sigma_{\uparrow\uparrow}$ scattering may be understood in the following way.

\subsubsection{Divergence of scattering cross section at $k\rightarrow 0$}
From Fig. \ref{Fig5}(a), we see that the total scattering cross section diverges at $k \rightarrow 0$, where $\hbar^2 k^2/ (2m_e) = E$ for the spin up channel. This may be understood from a partial wave analysis of scattering of a spinless particle from a two-dimensional hard disk. In terms of the partial wave phase shifts $\delta_l$, the total scattering cross section is given by the expression\cite{Lapidus}
\begin{equation}
\sigma_{\uparrow \uparrow} = \frac {4} {k} \big( \sin^2 \delta_0  +  2 \sum_{l=1}^{\infty}   \sin^2{\delta_l}   \big),
     \label{Eq:total_cross_partial}
\end{equation}
where $l$ is the angular momentum channel. For small $k$,  the spin up particle sees essentially an infinite potential barrier inside the skyrmion (Fig. \ref{Fig4}), which therefore behaves as a hard disk. For a hard disk of radius $a$, $\tan{\delta_l} = J_l(ka)/Y_l(ka)$, where $J_l(ka)$ and $Y_l(ka)$ are the Bessel and Neumann functions. 
In the limit $k \rightarrow 0$, the phase shifts are 
$\delta_0 \approx  \frac{\pi}{2} (\gamma + \ln{\frac{ka}{2}})^{-1} $, while 
$\delta_l \propto (ka)^{2l}$ for $l \ge 1$. Plugging these results into Eq. (\ref{Eq:total_cross_partial}), we immediately see that the $s$ wave channel dominates, so that $\sigma = 4  \delta^2_0 / k 
\sim \pi^2  (k \ln^2{ka})^{-1}$, which diverges in the small $k$ limit. 
The same divergence occurs for the spin down channel $\sigma_{\downarrow \downarrow}$ at $E = J$, corresponding to $k \rightarrow 0$ for the spin down case, as seen from Fig. \ref{Fig5} (b). The same divergence is also seen from Fig. \ref{Fig6} (b) for a model potential discussed in the next subsection.

\subsubsection { Peaked structure at $E = J$}

\begin{figure}[tb!]
    \centering
   \includegraphics[width=1\linewidth]{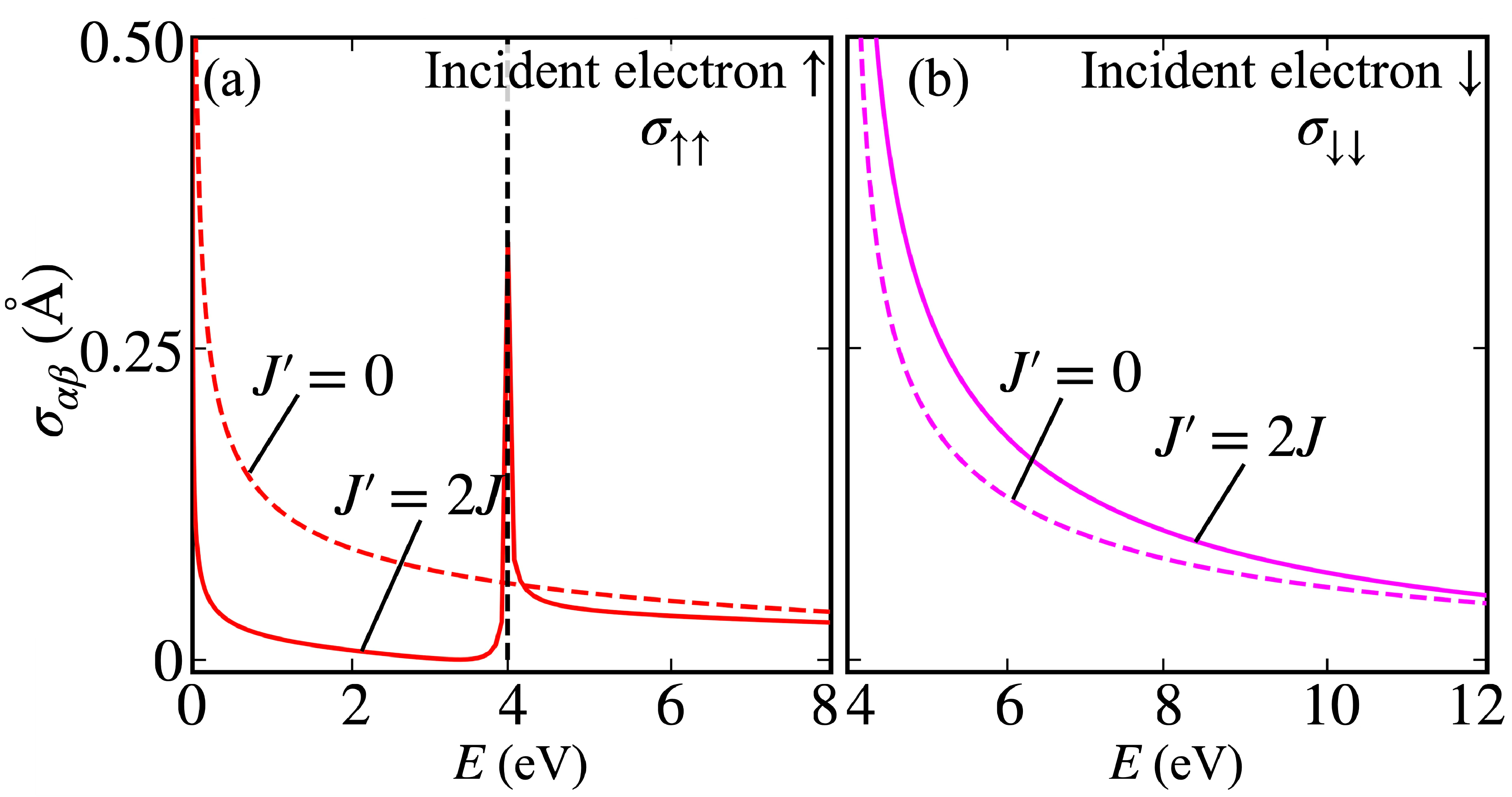}
    \caption{ (a) Scattering cross section   $\sigma_{\uparrow \uparrow}$ from the model potential, Eq. (\ref{model-V}), 
    for two different spin-flip potential strengths $J'$, to illustrate the origin of the logarithmic divergence at $k = 0$ and the peak at $E = J$ ($ J = 4$ eV). The latter originates due to the scattering into the propagating spin down channel beyond $E > J$, where there is a log divergence at $E = J$ for the spin down channel, analogous to the divergence at $k =0$ for the spin up channel.  Evanescent spin down states exist for $E < J$, causing the peak structure to extend below $J$.    
    (b) Scattering cross section $\sigma_{\downarrow \downarrow}$ for the spin down channel showing divergence at $E = J$.
   The spin-flip scattering cross sections $\sigma_{-\beta, \beta}$, which are non-zero for $E > J$ are not shown.  }
    \label{Fig6}
\end{figure}

The peaked structure at $E = J$, seen from Fig. \ref{Fig5}(a) for $\sigma_{\uparrow \uparrow}$, arises due to the divergence in the spin down channel $\sigma_{\downarrow \downarrow}$,
which is coupled to the scattering of the spin up electrons due to the spin-flip term in the scattering potential. To illustrate this, we take the following model potential, where we can turn the spin-flip potential $J'$ on or off. We consider a square-well type potential, where $J, J'$ have a fixed value 
for $r < a_0$ and zero otherwise, so that
\begin{equation}
    V(\bs r)  
         = \begin{pmatrix}
        J & J' \\
        J' & -J
         \end{pmatrix}.
         \label{model-V}
\end{equation}
The square-well radius $a_0$ can be taken sufficiently small, if one wishes to suppress other scattering features such as the Ramsauer-Townsend dips discussed in the next subsection. The results are  shown in Fig. \ref{Fig6}. When the spin-flip term $J' = 0$, we see the usual divergence at $E = 0$ in $\sigma_{\uparrow\uparrow}$, while with the inclusion of the spin-flip term $J' \ne 0$, we get the additional peak at $E = J$ corresponding to the $k =0$ divergence in the spin-down channel, which now mixes up with the spin up states. Furthermore, the divergence is broadened due to the existence of the evanescent states for the spin-down electrons for $E < J$, which exist in the neighborhood the skyrmion but do not have sufficient energy to exist far away from the scattering center. This was elaborated already in the discussions leading to Eq. (\ref{Eq:G0-limit2}).

\subsubsection { Ramsauer-Townsend dips near $E = J$} 

\begin{figure}[tb!]
    \centering
   \includegraphics[width=0.7\linewidth]{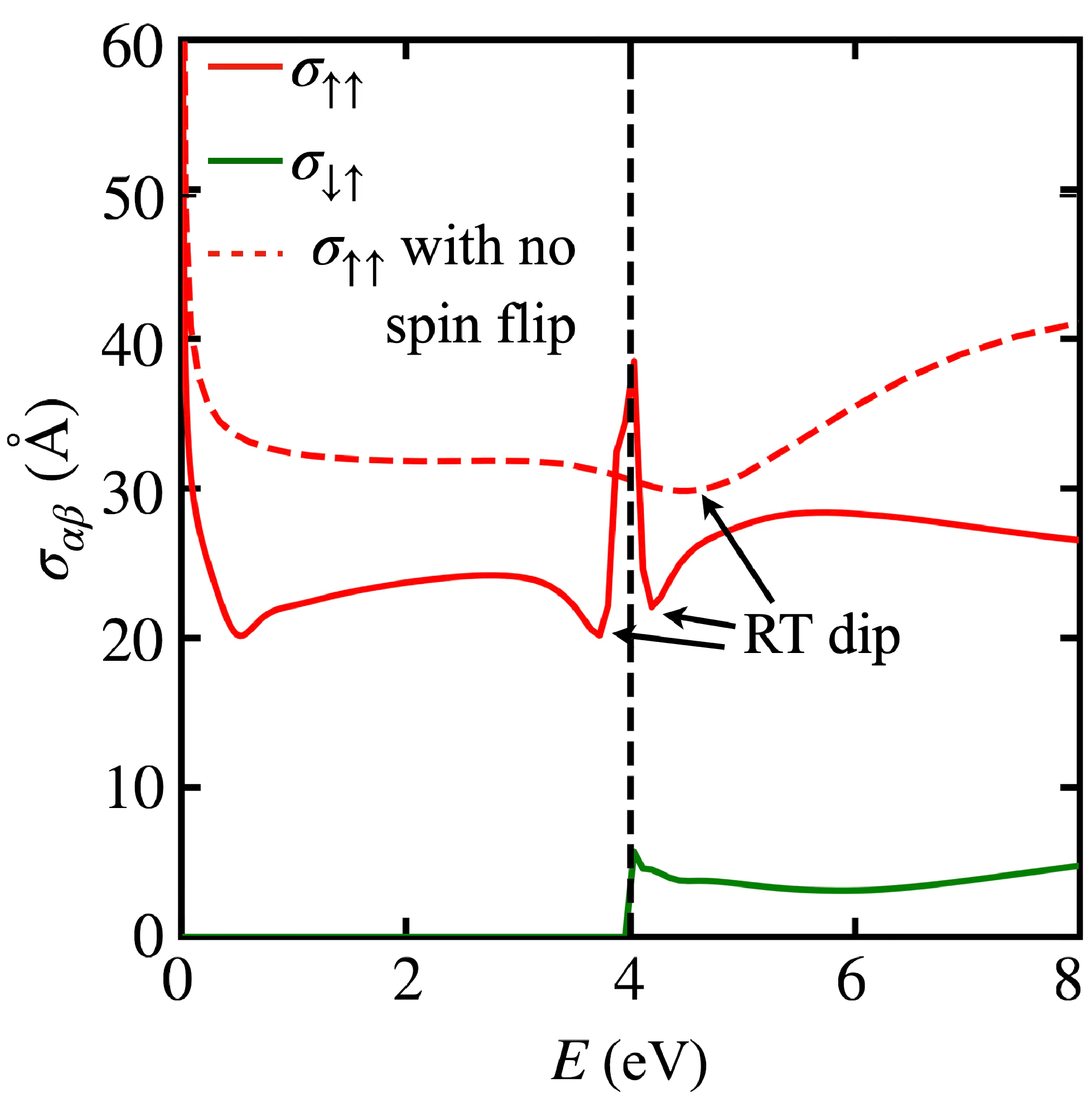}
    \caption{
 Total scattering cross sections (solid lines) for an incident spin-up electron with energy $E$. 
With the spin-flip skyrmion potential, Eq. (\ref{Eq:potential}), set equal to zero, $V_{\uparrow\downarrow} = V_{\downarrow\uparrow} = 0$, a Ramsauer-Townsend (RT) dip is seen around $E \approx 4.4 $ eV (red dashed line). With the spin-flip potential included, the Ramsauer-Townsend dip shifts in energy and becomes integrated with the peaked structure at $E = J = 4 $ eV. 
}
    \label{Fig7}
\end{figure}

In addition to the divergent peaks at $E = 0$ and $J$, there exists a dip in the total scattering cross section for the incident energy slightly greater than $J$, when there is no  spin-flip potential. When the spin-flip potential is included, the dip merges with the divergent peak at $E = J$, as indicated in Fig. \ref{Fig7}.
This may be interpreted in terms of the Ramsauer-Townsend  effect\cite{Ramsaur_1921, Townsend_1922, RT-expt, Rashba-Ramsauer}, first observed in the low-energy electron scattering from noble gases, which shows an unexpected high transmission or a sharp minimum in the scattering cross section. 

\begin{figure}[tb!]
    \centering
\includegraphics[width=0.8\linewidth]{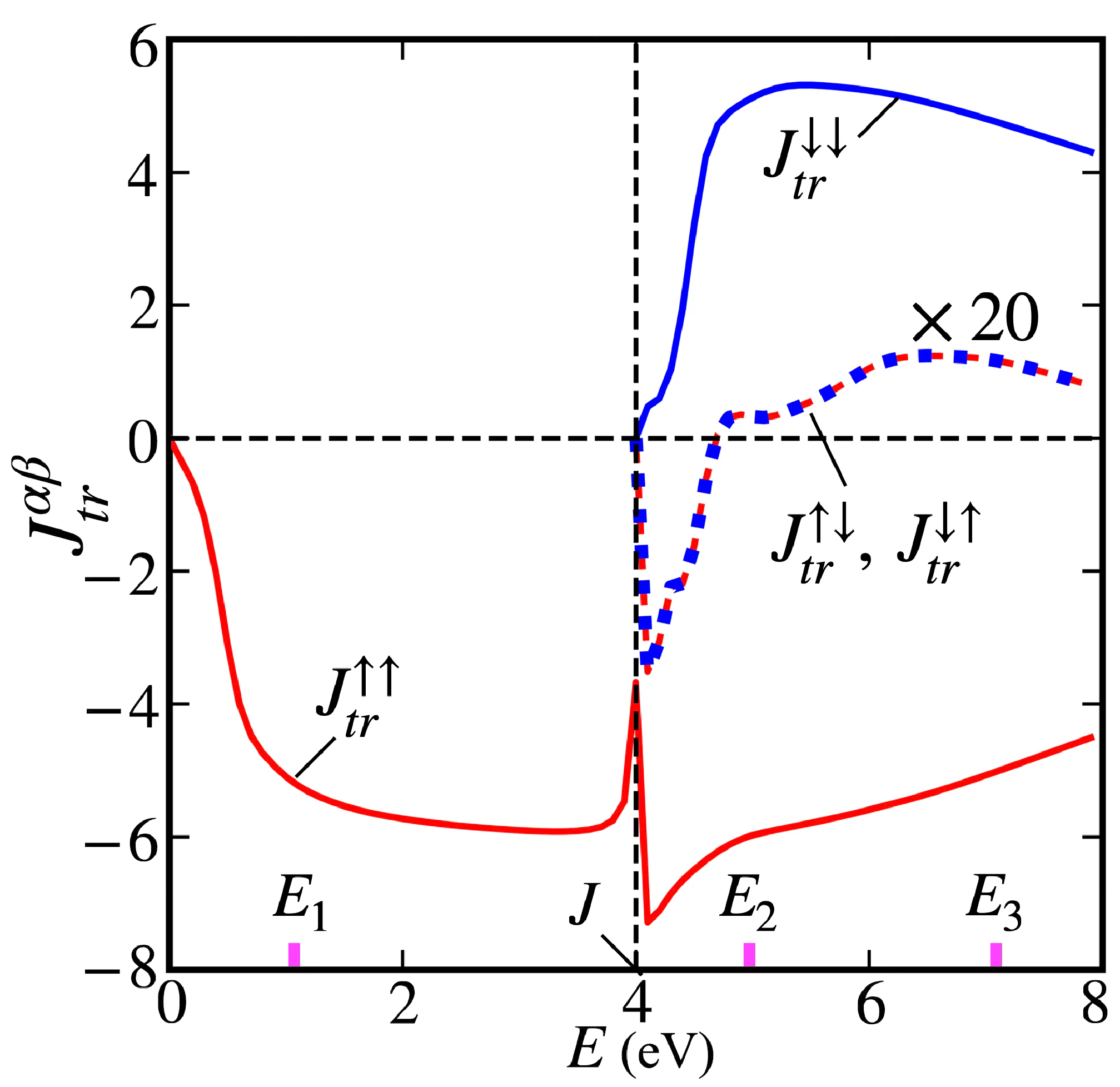}
    \caption{
 Spin resolved transverse current $J_{tr}^{\alpha \beta}$ as a function of the incident energy $E$, obtained from Eq. (\ref{Eq20}). 
 Spin up electrons scatter predominantly to the right ($J_{tr}^{\uparrow \uparrow} < 0$),
 while spin down electrons scatter predominantly to the left  ($J_{tr}^{\downarrow \downarrow} > 0$).
 For $E < J$, only $J_{tr}^{\uparrow \uparrow}$ is non-zero since spin down electrons do not have enough energy to exist outside the skyrmion as a propagating state.  
 The different types of scattering for the energies marked on the $x$-axis are sketched   in Fig. \ref{Fig10}. 
}
    \label{Fig8}
\end{figure}

\begin{figure}[hbt!]
    \centering
   \includegraphics[width=.9\linewidth]{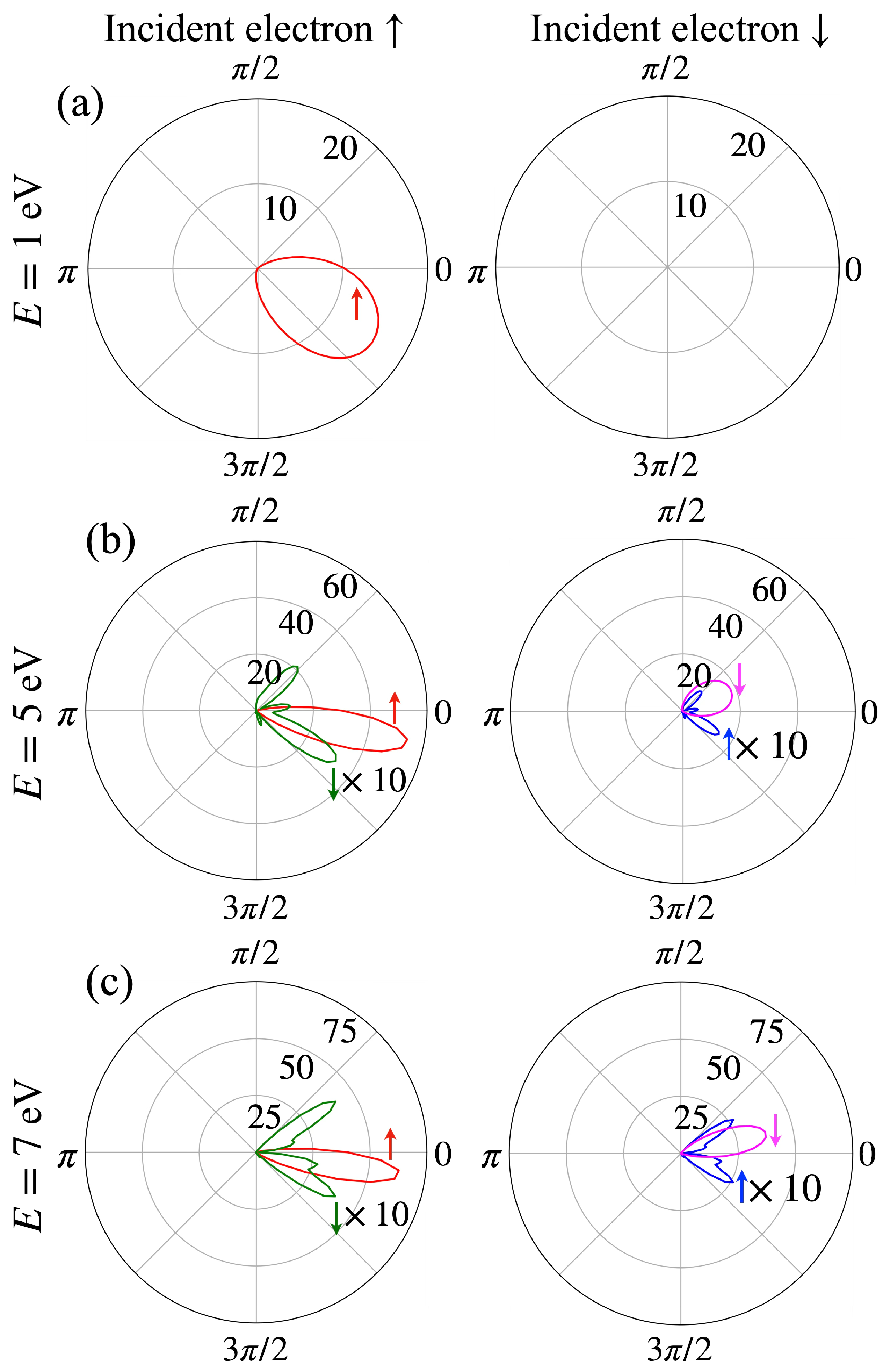}
    \caption{
 Differential scattering cross section $|f_{\alpha \beta} (\theta)|^2$   for three representative energies, marked on the $x$-axis of Fig. \ref{Fig8}: (a) $E_1 = 1$ eV, (b) $E_2 = 5$ eV, and (c) $E_3 = 7$ eV. The left (right) column corresponds to the incident spin up (down) electron. The spin of the scattered electron is indicated by the arrows inside each figure. The right column, top is empty because there is no incident spin down electron possible for $E < J = 4$  eV. The three scattering scenarios corresponding to the three different energy regimes are shown schematically in Fig. \ref{Fig10}.
}
    \label{Fig9}
\end{figure}

\begin{figure}[hbt!]
    \centering
\includegraphics[width=0.9\linewidth]{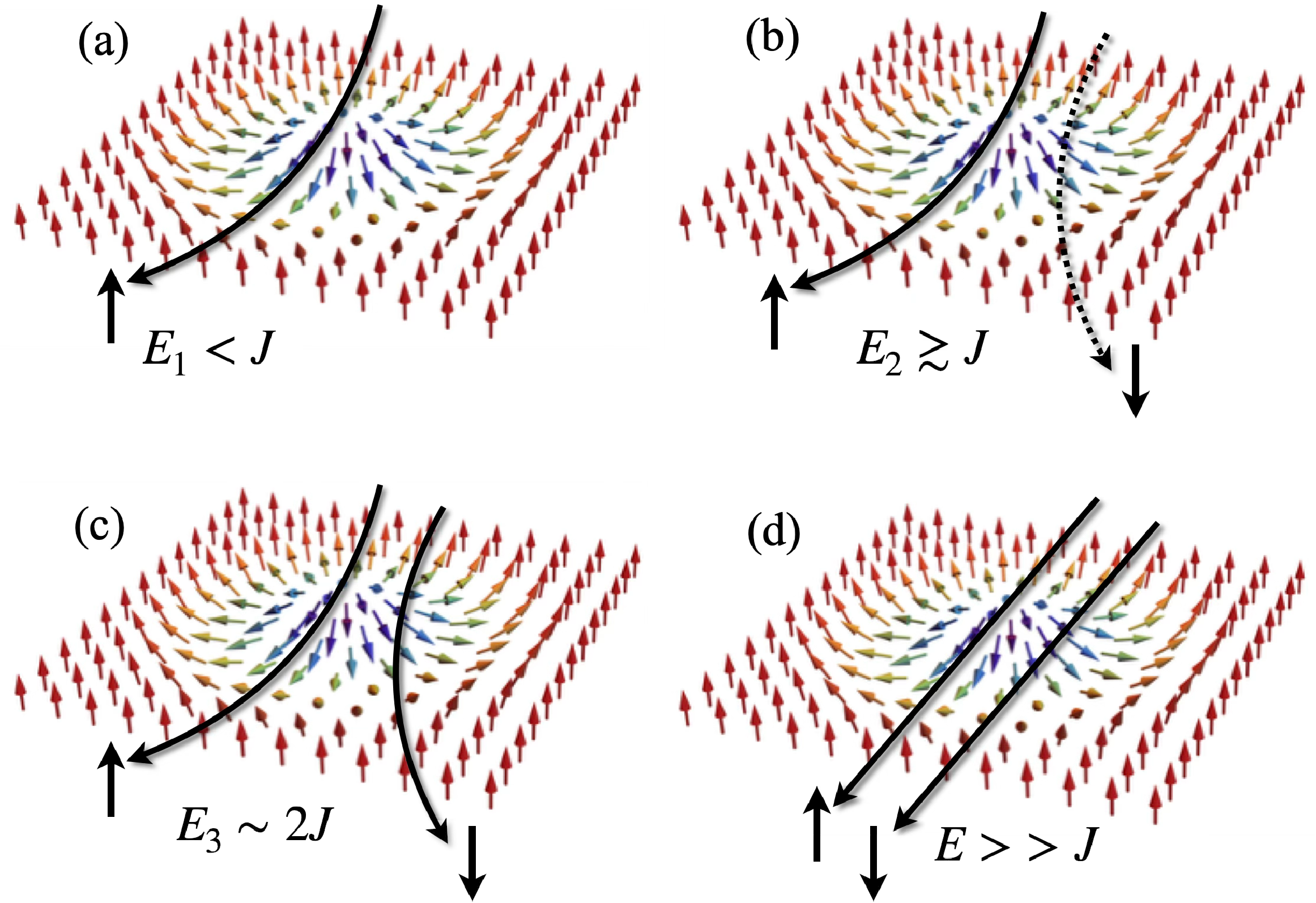}
    \caption{
   Electron scattering in  different energy regimes. 
   Scattering in the spin-preserved channel only is shown, and the small spin-flip scattering (see Fig. \ref{Fig8}) is omitted for clarity.
 (a) $E_1 < J$: Here, only spin up electrons can exist outside the skyrmion as propagating states, leading to the result $\sigma^H = \sigma^{SH}$, Eq. (\ref{Eq20}).
 (b) $E_2 \gtrsim J$: 
 There is a prominent spin up current, but somewhat weaker spin down current (indicated by the dashed line).
 (c) Intermediate energy, $E_3 \sim 2 J$: Both spin up and down currents are prominent, and nearly equal in magnitude, leading to $\sigma^H \rightarrow 0$ but a large $\sigma^{SH}$ (see Fig. \ref{Fig11}). (d) Very high energy, $E \gg J$: Scattering is negligible for both spins, leading to $\sigma^H = \sigma^{SH} =0$. 
 Typical energies for these scenarios are marked on the $x$-axis in Figs. \ref{Fig8} and \ref{Fig11}. 
}
    \label{Fig10}
\end{figure}

The effect is illustrated by considering the 
transmission coefficient of a plane wave propagating through a 1D potential barrier with energy higher than the barrier height.  The text-book result is\cite{Zettlie}
\begin{equation}
    T = \left[1+ \frac{V^2}{4E (E-V)}\sin^2{(2\pi a/ \bar \lambda)} \right]^{-1},
    \label{Eq:T}
\end{equation}
where $ \bar \lambda = 2 \pi \hbar \ [ 2 m (E-V)]^{-1/2}$ is the particle wavelength inside the barrier, and $V$ and $a$ are the barrier height and width, respectively. 
As seen from Eq. (\ref{Eq:T}),
when  an integer number of half wavelengths fits exactly in the barrier length, $n  \bar \lambda / 2 = a$, $n =1, 2, 3, ...$, 
the particle does not scatter at all and there is perfect transparency ($T = 1$), which is physically interpreted to be due to the wave interference with the reflected component. 

In higher dimensions, such a matching cannot be done perfectly for the incoming planewave, but the remnants of the Ramsauer-Townsend transparency appears as a noticeable dip in the scattering cross section.
We may estimate the energy of the Ramsauer-Townsend dip for the simplest case, where there is no spin-flip scattering, so that it boils down to an effective spinless problem. The dip is shown by an arrow in Fig. \ref{Fig7} for the skyrmion potential Eq. (\ref{Eq:potential}).
When the spin-flip potential is set to zero, the dip occurs at $E \approx 4.4$ eV. Matching the half wavelength $\bar \lambda / 2$ of the electron inside the barrier, which we approximate as a uniform barrier of height $J$, to the skyrmion size, we get $\bar \lambda / 2 = a = 2 \lambda$. This yields the corresponding energy of the Ramsauer-Townsend dip to be 
$E = J + (2m_e)^{-1}\hbar^2 (2 \pi/\bar \lambda)^2 \approx 4.1$ eV.
The agreement with the numerical result is reasonable considering the severity of the approximations used. 

For the full scattering problem, which includes the spin-flip potential, the Ramsauer-Townsend dip shifts down in energy and merges with the peaked structure at $E = J$, forming a double well type structure in the scattering cross section as seen in Fig. \ref{Fig7}.
Deeper understanding of the origin of the Ramsauer-Townsend dip for the 2D case, if desired, may be obtained by considering the phase shifts of the various spin and angular momentum channels.

\section{ Topological Hall and spin Hall conductivities}

The asymmetry in the scattering cross section, already indicated from Figs. \ref{Fig2} and \ref{Fig5}, results in a net charge and/or spin current flowing in the transverse direction,
leading to the charge or spin Hall effect.
We consider scattering of electrons with a specific energy $E$, with equal spin populations when $E > J$. For $E < J$, only spin up electrons can propagate both in the incident beam as well as in the scattered beam, so that the incident beam consists of the spin up electrons only. 

The Hall conductivity is determined by the asymmetry in the electron scattering, which produces an asymmetric current to the left vs. the right of the scatterer. 
Denoting the excess transverse current to the left over the right of the scatterer as $J_{tr}^{\alpha \beta}$ for a single spin channel, with incident spin $\beta$ and scattered spin $\alpha$, the transverse current is easily found to be (see Appendix)
\be
J_{tr}^{\alpha \beta}  =   \frac{\hbar k_\alpha} {m_e}  \int_0^{2 \pi}  |f_{\alpha\beta}(\theta)|^2 \sin \theta\ d\theta.
\label{Eq19}
\ee

Since the incoming current may be taken as proportional to the applied electric field, $J_{in} \propto k \propto E$ (in the relaxation time approximation for a metal: $\hbar k / m = eE \tau$), we define the Hall conductivity by simply the ratio of the transverse current to the incident current, taking the material-dependent prefactor to be unity. For the computation of the Hall conductivity, there are two energy regimes.
A quick calculation  yields (see Appendix):
\begin{eqnarray}
\sigma^H &=& \sigma^{SH} =  \int_0^{2 \pi}  |f_{\uparrow \uparrow}|^2 \sin \theta\ d\theta \ \ \ \ \ ( E < J), \nonumber \\
 \sigma^{H/SH}  &=& \sum_{\alpha\beta} \xi_\alpha \ \bar k_\alpha  \int_0^{2 \pi}  |f_{\alpha\beta}|^2 \sin \theta\ d\theta  \ ( E > J),
\label{Eq20}
\end{eqnarray}
where the summation is over the two spins, 
$\bar k_\alpha = (k_\uparrow + k_\downarrow)^{-1} k_\alpha$,  and the prefactor    $\xi_\alpha = 1$ for $\sigma^H$, while $\xi_\uparrow =-\xi_\downarrow = 1$ for $\sigma^{SH}$.
For  $E > J$, both spin up and down propagating states are possible, and, as already stated, we have assumed an equal spin mixture for the incident beam. 

\begin{figure}[tb!]
    \centering
\includegraphics[width=0.8\linewidth]{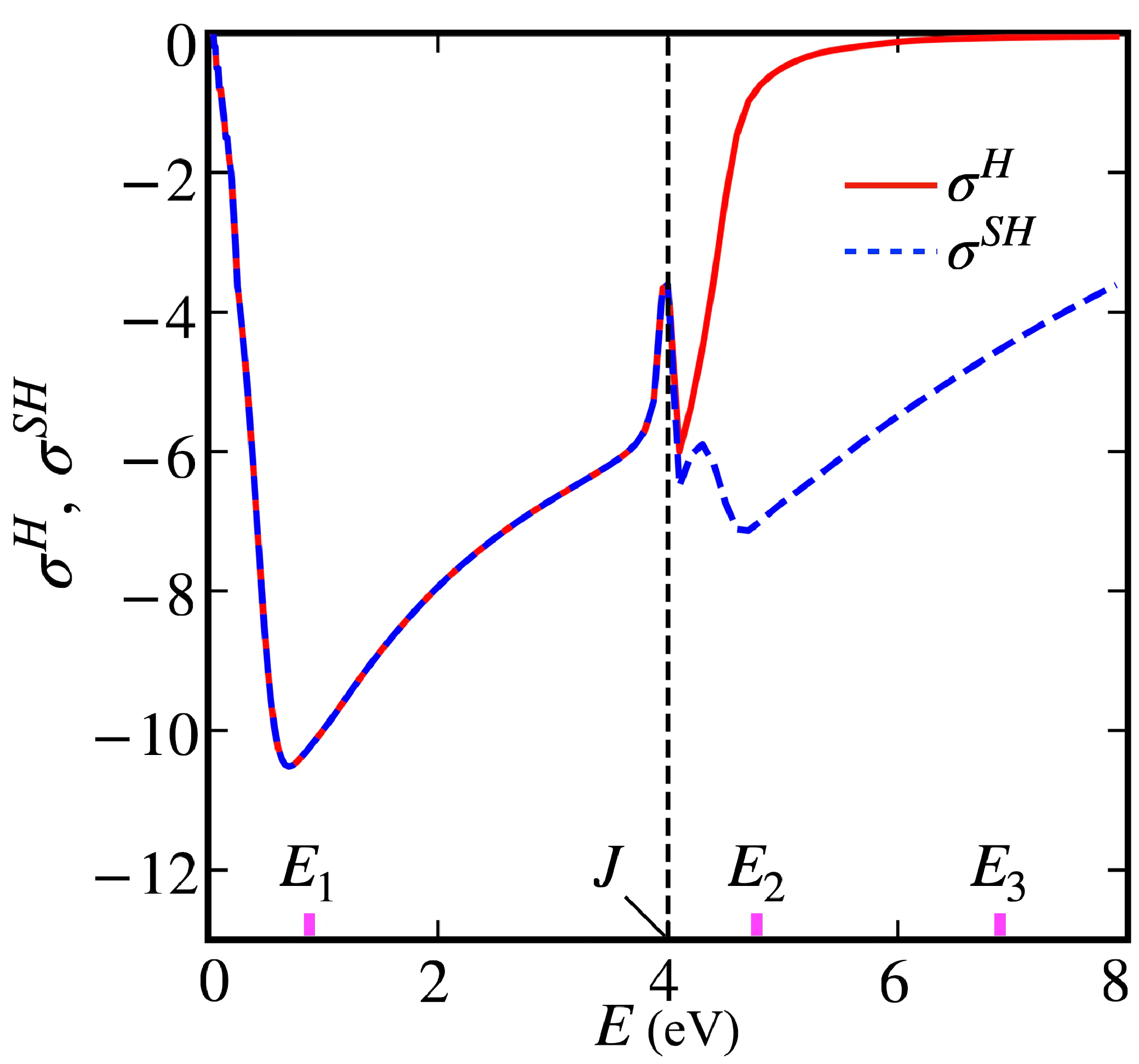}
    \caption{
 Topological Hall conductivities, $\sigma^{H}$ and $\sigma^{SH}$, calculated from Eq. (\ref{Eq20}), as a function of the energy of the incident electron.    
}
    \label{Fig11}
\end{figure}

The spin-resolved transverse current  $J_{tr}^{\alpha\beta}$ computed from Eq. (\ref{Eq19}) is shown in Fig. \ref{Fig8} as a function of the incident energy $E$. 
The differential scattering cross section for three representative energies, indicated on the $x$-axis in Fig. \ref{Fig8},  are shown in Fig. \ref{Fig9}. There are several features to note for the scattering cross section and the transverse current.

(i) First of all, an important point to note is that the spin up electrons scatter predominantly to the right ($J_{tr}^{\uparrow\uparrow} <0 $), while spin down electrons scatter predominantly to the left ($J_{tr}^{\downarrow\downarrow} > 0 $)   due to the chiral spin texture of the skyrmion. If the spin texture is changed from skyrmion (winding number $m = 1$)  to anti-skyrmion type ($m = -1$), the opposite would happen.

(ii) For $E < J$, only spin up electrons can exist in the incident and scattered beams, so that only $J_{tr}^{\uparrow \uparrow}$
is non-zero in this energy range. For $E > J$,  both spin up and down electrons can exist in the incident and scattered beams, so that all transverse currents are non-zero.

(iii) The sharp peaked structure at $E \approx J$ comes from the divergence in the scattering cross section in the spin down channel plus the evanescent waves for $E \lesssim J$ created in the skyrmion region in the same channel as discussed in  Section \ref{IIA}.

(iv) The spin flip currents are equal in magnitude, $J_{tr}^{\uparrow \downarrow} = J_{tr}^{\downarrow \uparrow}$, which can be inferred from the  current expression Eq. (\ref{Eq19}) and the structure of the $T$-matrix, Eq. (\ref{Eq:f}), using the hermiticity of the Hamiltonian.
Furthermore, they are much smaller than the spin preserved currents $J_{tr}^{\sigma \sigma}$.

(v) In the high-energy limit $E \gg J$, the electrons in essence don't see the scattering center, so that all currents eventually go to zero (not shown in Fig. \ref{Fig8}).

The distinctive behavior of the transverse current in the different energy regions is schematically shown in Fig. \ref{Fig10}.

The computed topological Hall and spin Hall conductivities, $\sigma^H$ and $\sigma^{SH}$, respectively, are shown in Fig. \ref{Fig11} as a function of the incident electron energy. 
The conductivities  can be understood from the scattering cross section and the transverse currents just discussed.
They show four distinct behaviors, as sketched in Fig. \ref{Fig10}:

(a)  When $E \le J$, only the spin up electrons can exist outside the skyrmion as propagating waves, so that for this energy, $\sigma^H = \sigma^{SH}$.

(b) When $E \gtrsim J$ (for example, $E_2$ in Fig. \ref{Fig11}), both spin states are allowed as propagating states, but the scattering cross section 
for the spin up electron is stronger than the same for the spin down electron (see Fig. \ref{Fig8}), which produces a large $\sigma^{SH}$ but a somewhat smaller $\sigma^H$.

(c) For intermediate energies, $E \sim 2J$, both spin states in the incident wave  are scattered comparatively by the same amount, so that $\sigma^H \approx 0$, but there is a strong $\sigma^{SH}$ as seen for $E_3$ in Fig. \ref{Fig11}.

(d) Finally, in the trivial case of $E \gg J$, the electrons in essence don't see the scattering center, so that there is virtually no scattering, and $\sigma^H = \sigma^{SH} = 0$. 

\section {skyrmions with higher winding number: Effect of Landau levels}

\begin{figure*}[hbt!]
    \centering
   \includegraphics[width=0.8\linewidth]{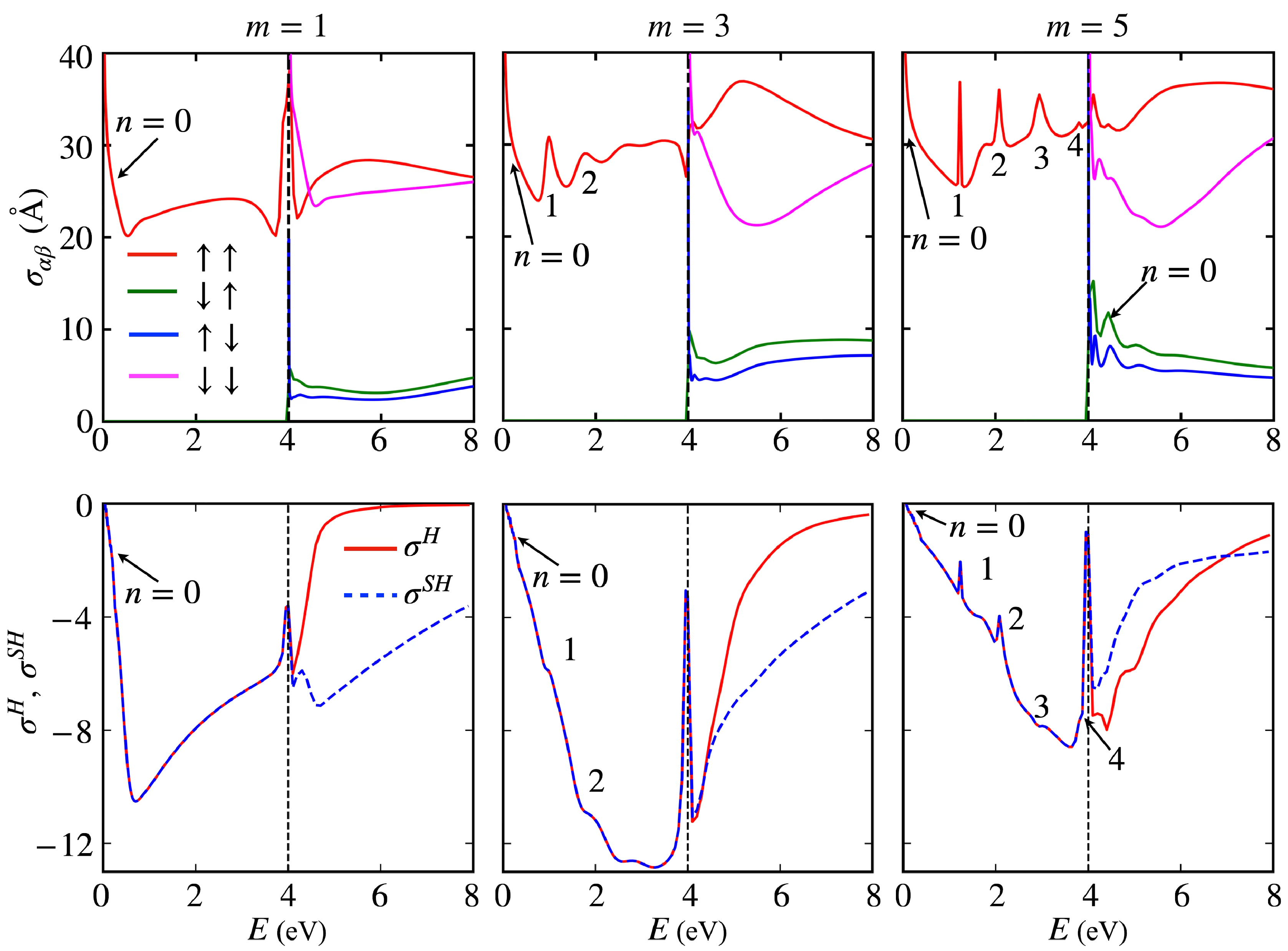}
    \caption{ {\it Top panel}, Spin-dependent scattering cross sections for skyrmions with winding numbers $m = 1$, $3$, and $5$. Spin values $\alpha, \beta$ are color coded as indicated in the {\it top left} figure. {\it Bottom panel}, The corresponding topological charge Hall and spin Hall conductivities, $\sigma^H$ and $\sigma^{SH}$, respectively.
    With increasing $m$, the magnetic field becomes progressively stronger, resulting in several Landau levels that can be accommodated inside the skyrmion. 
    The numbers inside the figures indicate resonant scattering peaks due to various Landau levels, labeled by $n$, due to the effective magnetic field as discussed in the text.  The resonant peaks show up for $m \ge 2$. The resonance scattering effect of the lowest Landau level peaks, $n = 0$, is washed out as the energy falls in the divergent cross section region close to $E = 0$. 
    }
    \label{Fig12}
\end{figure*}

\begin{table}
\begin{center}
\caption{Energies $E_n$ and radii  $r_n$ of Landau levels for skyrmions with different winding number $m$, for skyrmion size $\lambda=10 $ \AA. Only the Landau levels with radius $r_n < \lambda$ are listed for each $m$. The last column lists the peak positions
in the calculated $\sigma_{\uparrow\uparrow}$ (see Fig. \ref{Fig12} for $m$ = 1, 3, and 5).
}
\setlength{\tabcolsep}{3pt}
\renewcommand{\arraystretch}{1.3}
\begin{tabular}{c| c c c| c}
\hline
\hline
Winding & \multicolumn{3}{c|}{Landau level} & Peak positions  \\
number, m & n & $r_n$(\AA) & $E_n$(eV) & in $\sigma_{\uparrow\uparrow}$ (eV)  \\
\hline
\multirow{ 1}{*}{1} & 0 & 7.1 & 0.08 & -\\
\hline
\multirow{ 2}{*}{2} & 0 & 5.0 & 0.15 & -\\
& 1 & 8.7 & 0.45 & 0.91\\
\hline
\multirow{ 3}{*}{3} & 0 & 4.1 & 0.23 & -\\
& 1& 7.1 & 0.68 & 0.98\\
& 2 & 9.1 & 1.13 & 1.75\\
\hline
\multirow{ 4}{*}{4} & 0 & 3.5 & 0.30 & -\\
& 1 & 6.1 & 0.90 & 1.10\\
& 2 & 7.9 & 1.50 & 1.91\\
& 3 & 9.3 & 2.10 & 2.81\\
\hline
\multirow{ 5}{*}{5} & 0 & 3.2 & 0.38 & -\\
& 1 & 5.5 & 1.13 & 1.23\\
& 2 & 7.1 & 1.88 & 2.08\\
& 3 & 8.4 & 2.63 & 2.94\\
& 4 & 9.5 & 3.38 & 3.80\\
\hline
\hline
\end{tabular}
\label{Table1}
\end{center}
\end{table}

Due to the spin texture of the skyrmion,
the electron experiences an emergent magnetic field,
given by\cite{nagaosa-review,Mandal} 
\begin{equation}
    B(r) = \frac{m \Phi_0}{2\pi} \times  \frac{  \pi}{2\lambda r}\sin\frac{\pi r}{\lambda},
    \label{Eq21}
\end{equation}
where $\Phi_0 = h/e$ is the flux quantum, and $m$ is the winding number. 
The presence of a magnetic field leads to the Landau levels and the associated edge currents. The latter will die out in the long distance limit, but the presence of the Landau levels will affect the scattering process via resonance scattering. 
Even though the magnetic field expression Eq. (\ref{Eq21}) is strictly valid in the adiabatic limit $J \rightarrow \infty$, so that the itinerant electron spin adiabatically follows the local spin alignment, the adiabatic parameter corresponding to the Landau-level energies $E_n$ is sufficiently large that our analysis in terms of the magnetic field is reasonable.

In order to examine the effect of the Landau levels, we estimate the size and the energies, by simply taking the average  magnetic field inside the skyrmion, $ B = m \Phi_0/\pi \lambda^2$, where $\Phi_0$ is the flux quantum.
The energy of the Landau levels are given by $E_n = (n + \frac{1}{2}) \hbar \omega_B$ with the radius given by $r_n = l_0 (2n+1)^{1/2}$, where  $\omega_B$ is the cyclotron frequency and the magnetic length $l_0 = (eB/\hbar)^{-1/2}  = \lambda \sqrt{2m_e}$ for the skyrmion. 
Although the finite size of the skyrmion is expected to modify the Landau-level energies, we neglect such corrections because they lie beyond the accuracy of the present order-of-magnitude analysis.
Furthermore, when the Landau-level radius exceeds the skyrmion radius, the corresponding levels can no longer be confined within the skyrmion and therefore do not form bound states. 
An additional point to note is that the highly-degenerate Landau levels 
for the 2D plane are no longer degenerate due to the finite size of the skyrmion, but will somewhat spread out in energy. The degeneracy of each Landau level is estimated from the standard expression, $D = BA/\Phi_0 = m$, $A=\pi \lambda^2$ being the area of the skyrmion. This will produce a broadening of the resonance scattering peaks.

With these approximations, the first few energies of the Landau Levels and the corresponding radii for skyrmions with different winding numbers are listed in Table \ref{Table1}, where only those Landau levels, whose radius is less than the skyrmion radius, are included. 
These bound states produce scattering resonances seen in Fig. \ref{Fig12} for the winding numbers $m = 3$ and $5$, which are very prominent for the $\sigma_{\uparrow \uparrow}$ channel. The resonance peak positions are listed on the last column in Table \ref{Table1}.
The resonance peak positions can be identified with the Landau-levels, with the energies matching quite well for larger winding numbers, e. g., $m =5$. For lower $m$, as seen from Table \ref{Table1}, the Landau-level energies can differ from the energy of the peak positions substantially. This is due to the crude approximation of using the uniform field in calculating the Landau-level energies. 
Note that the lowest Landau level ($ n =0$) has a low enough energy that it does not show up as a scattering peak for all $m$, since it merges with the  divergent cross section at $E = 0$. However, the lowest Landau level does have a noticeable signature for $E > J$ as indicated by an arrow for $m = 5$ in the top panel for $\sigma_{\downarrow \uparrow}$. 

The corresponding topological Hall conductivities are shown in the bottom panel of Fig. \ref{Fig12}, where $\sigma^H$ and $\sigma^{SH}$ for higher winding numbers ($m = 3$ and 5) are compared with $m = 1$. Increasing $m$ makes the peak structure at $E = J$ more prominent. For skyrmions with higher winding numbers ($m =5$), the Landau-level peak structure of the scattering cross section carries over to the Hall conductivities as well, while for $m =3$, the sharp peaks are absent, but the signatures of the Landau levels appear as plateaus in the Hall conductivity.

\section{Conclusion}

In conclusion, we have studied electron scattering from a single skyrmion using the full Green’s function and the Lippmann–Schwinger equation, a nonperturbative formulation valid for arbitrary scattering strength. The dimensionless scattering strength is characterized by the ratio of the exchange interaction $J$ to the electron energy $E$. In our calculations, we typically fixed $J$ and varied the electron energy to tune $J/E$.
The Green’s function and the $T$-matrix were obtained numerically through real-space matrix inversion. In the weak-coupling limit, our method correctly reproduces the results of the Born approximation, providing an important validation of the approach. With the change of the scattering strength, several notable features emerge in the scattering cross section that are absent in the weak-scattering regime, including intermediate-coupling resonances, Ramsauer–Townsend dips, and pronounced Landau-level resonances for skyrmions with higher winding numbers.
These features strongly influence the topological and spin Hall conductivities, which exhibit distinct behavior across different energy regimes. In particular, our results suggest that small changes in the incident electron energy can lead to significant tuning of the Hall conductivities, especially near resonance peaks.
Although the present work focuses on electron scattering from skyrmions, the formalism is readily applicable to other noncollinear spin textures of current interest.

\section{Acknowledgements} 
BRKN acknowledges HPCE, IIT Madras for providing the computational facilities. AM thanks MoE India for the PMRF fellowship. SS
thanks SERB India for the VAJRA fellowship and IIT Madras for the IIE fellowship.

\appendix
\section{Transverse current and topological Hall conductivity}
\label{appenA}

The quantum mechanical current density is given by the expression
\be
\bs j  = \frac{\hbar}{m_e}  \ \text{Im}\ (\psi^{\ast} \bs \nabla \psi).
\label{eq1}
\ee
Consider the electron scattering into one spin channel, incident spin $\beta$ and
outgoing spin $\alpha$, so that the wave function after scattering in the 2D problem is
\be
\psi (\bs r) = e^{i k_\beta x} + f_{\alpha\beta}(\theta) \frac{e^{i k_\alpha r} } {\sqrt r},
\label{eq2}
\ee
where the electron is incident along $\hat x$. 
The incident current density using Eq. (\ref{eq1}) is given by
\be
\bs j_{in} = J_{in} \ \hat x = \frac{\hbar k_\beta} {m_e} \hat x.
\ee
The scattered current is in the radial direction, which follows from Eq. (\ref{eq2}) to be 
\be
\bs j_{sc} = \frac{\hbar k_\alpha} {m_e} |f_{\alpha\beta}(\theta)|^2 \frac {\hat r} {r}
\ee
in the large distance $ r \rightarrow \infty$ limit.
The radial current falls off with distance $r$ as usual, but the scattered current per unit angle (which we call $J_{sc}$) is independent of the radial distance:
\be
J_{sc} = \frac{\hbar k_\alpha} {m_e} |f_{\alpha\beta}(\theta)|^2 .
\ee
The total transverse current along $\hat y$ is obtained by multiplying with the $\sin \theta$ factor and integrating
\be
J_{tr}  =   \frac{\hbar k_\alpha} {m_e}  \int_0^{2 \pi}  |f_{\alpha\beta}(\theta)|^2 \sin \theta\ d\theta.
\ee

Since the incoming current may be taken as proportional to the applied electric field, $J_{in} \propto k \propto E_x$ (in the relaxation time approximation for a metal: $\hbar k / m = eE \tau$), we define the Hall conductivity by simply the ratio of the transverse current to the incident current, taking the material-dependent prefactor to be unity:
\be
\sigma_{\alpha\beta}^H = J_{tr}/ J_{in}   =  \frac{ k_\alpha} {k_\beta} \int_0^{2 \pi}  |f_{\alpha\beta}(\theta)|^2 \sin \theta\ d\theta.
\label{eq7}
\ee

This equation is valid for a single spin channel. It can be generalized easily when both spin channels are present.
We simply have to compute the total transverse current and divide by the total incoming current taking into account both spins.
There are two different energy regimes for the computation of the Hall conductivity.

{\it Case 1: $ E < J$}. In this case, only spin up electrons can be present in both the incident and the scattered beams, and the spin down state can exist only in the skyrmion region. For this case, using Eq. (\ref{eq7}), we immediately find
\be
\sigma^H = \sigma^{SH} =  \int_0^{2 \pi}  |f_{\uparrow \uparrow}|^2 \sin \theta\ d\theta.
\label{eq8}
\ee

{\it Case 2: $ E > J$}. Here, both spin up and spin down states can exist both in the incident beam and the scattered beam. We have considered in the main text an equal spin mixture for the incident beam to compute the Hall conductivity, viz.,
\be
|\psi_{in} \rangle = \frac{1}{\sqrt{2}}\begin{pmatrix}
e^{ik_{\uparrow}x}\\
e^{ik_{\downarrow}x}
\end{pmatrix}.
\label{eq9}
\ee
For this case, after a few lines of algebra, the results for the Hall conductivity become
\begin{eqnarray}
    \sigma^H &=& \sum_{\alpha\beta} \bar k_\alpha  \int_0^{2 \pi}  |f_{\alpha\beta}|^2 \sin \theta\ d\theta, \nonumber \\
    \sigma^{SH}  &=& \sum_{\alpha\beta} \xi_\alpha \ \bar k_\alpha  \int_0^{2 \pi}  |f_{\alpha\beta}|^2 \sin \theta\ d\theta,
    \label{eq10}
\end{eqnarray}
where  $\xi_\uparrow = 1$, $\xi_\downarrow = -1$,  $\bar k_\alpha = (k_\uparrow + k_\downarrow)^{-1} k_\alpha$, and the summations are over the two spins.
If the incident beam has a different mixture of spin up and spin down components rather than the equal mixture Eq. (\ref{eq9}), then the Hall conductivity expressions Eq. (\ref{eq10}) can be easily generalized.

\bibliography{ref.bib}

@article{Bogdanov1,
author={Bogdanov, A. N. and Yablonskii, D. A.},
title={Thermodynamically stable ``vortices'' in magnetically ordered crystals. The mixed
state of magnets},
journal={Soviet Phys. JETP},
year={1989},
volume={68},
pages={101},
url={http://www.jetp.ras.ru/cgi-bin/dn/e_068_01_0101.pdf}
}

@article{bogdanov2,
title={ Theory of magnetic vortices in easy-axis ferromagnets},
journal = {Soviet Phys. Solid State},
volume = {31},
number = {99},
year = {1989},
author = {Bogdanov, A. and Kudinov, M.V. and Yablonskii, D.}
}

@Article{skyrmion1,
author={R{\"o}{\ss}ler, U. K.
and Bogdanov, A. N.
and Pfleiderer, C.},
title={Spontaneous skyrmion ground states in magnetic metals},
journal={Nature},
year={2006},
month={Aug},
day={01},
volume={442},
number={7104},
pages={797-801},
issn={1476-4687},
doi={10.1038/nature05056},
url={https://doi.org/10.1038/nature05056}
}

@Article{skyrmion2,
title = {Real-space observation of a two-dimensional skyrmion crystal},
author={Yu, X. Z. and Onose, Y. and Kanazawa, N. and Park, J. H. and Han, J. H. and Matsui, Y. and Nagaosa, N. and Tokura, Y.},
journal={Nature},
year={2010},
month={Jun},
day={01},
volume={465},
number={7300},
pages={901-904},
doi={10.1038/nature09124},
url={https://doi.org/10.1038/nature09124}
}

@Article{skyrmion3,
author={Yu, X. Z.
and Kanazawa, N.
and Onose, Y.
and Kimoto, K.
and Zhang, W. Z.
and Ishiwata, S.
and Matsui, Y.
and Tokura, Y.},
title={Near room-temperature formation of a skyrmion crystal in thin-films of the helimagnet {FeGe}},
journal={Nature Materials},
year={2011},
month={Feb},
day={01},
volume={10},
number={2},
pages={106-109},
issn={1476-4660},
doi={10.1038/nmat2916},
url={https://doi.org/10.1038/nmat2916}
}

@article{skyrmion4,
title={Thermodynamically stable magnetic vortex states in magnetic crystals},
journal = {Journal of Magnetism and Magnetic Materials},
volume = {138},
number = {3},
pages = {255-269},
year = {1994},
issn = {0304-8853},
doi = {https://doi.org/10.1016/0304-8853(94)90046-9},
url = {https://www.sciencedirect.com/science/article/pii/0304885394900469},
author = {A. Bogdanov and A. Hubert}
}

@article{Sadhan,
    author = {Adhikari, Sadhan K.},
    title = {Quantum scattering in two dimensions},
    journal = {American Journal of Physics},
    volume = {54},
    number = {4},
    pages = {362-367},
    year = {1986},
    month = {04},
    issn = {0002-9505},
    doi = {10.1119/1.14623},
    url = {https://doi.org/10.1119/1.14623}
}

@article{Morse-Feshbach,
    author = {Morse, P. M.  and Feshbach, H.},
    title = {Methods of Theoretical Physics},
    journal = {McGraw-Hill, New York},
    year = {1953}
}

@article{Zettlie,
    author = {Zettili, N.},
    title = { Quantum Mechanics Concepts and Applications (Second Edition)},
    journal = {Wiley},
    year = {2009}
}

@article{Denisov-prl,
  title = {Electron Scattering on a Magnetic Skyrmion in the Nonadiabatic Approximation},
  author = {Denisov, K. S. and Rozhansky, I. V. and Averkiev, N. S. and L\"ahderanta, E.},
  journal = {Phys. Rev. Lett.},
  volume = {117},
  issue = {2},
  pages = {027202},
  numpages = {5},
  year = {2016},
  month = {Jul},
  publisher = {American Physical Society},
  doi = {10.1103/PhysRevLett.117.027202},
  url = {https://link.aps.org/doi/10.1103/PhysRevLett.117.027202}
}

@Article{Denisov-sci-rep,
author={Denisov, K. S.
and Rozhansky, I. V.
and Averkiev, N. S.
and L{\"a}hderanta, E.},
title={A nontrivial crossover in topological {Hall} effect regimes},
journal={Scientific Reports},
year={2017},
month={Dec},
day={08},
volume={7},
number={1},
pages={17204},
issn={2045-2322},
doi={10.1038/s41598-017-16538-4},
url={https://doi.org/10.1038/s41598-017-16538-4}
}

@article{Denisov-prb-1,
  title = {General theory of the topological {Hall} effect in systems with chiral spin textures},
  author = {Denisov, K. S. and Rozhansky, I. V. and Averkiev, N. S. and L\"ahderanta, E.},
  journal = {Phys. Rev. B},
  volume = {98},
  issue = {19},
  pages = {195439},
  numpages = {12},
  year = {2018},
  month = {Nov},
  publisher = {American Physical Society},
  doi = {10.1103/PhysRevB.98.195439},
  url = {https://link.aps.org/doi/10.1103/PhysRevB.98.195439}
}

@article{Denisov-prb-2,
  title = {Topological {Hall} effect for electron scattering on nanoscale skyrmions in external magnetic field},
  author = {Denisov, K. S. and Rozhansky, I. V. and Potkina, M. N. and Lobanov, I. S. and L\"ahderanta, E. and Uzdin, V. M.},
  journal = {Phys. Rev. B},
  volume = {98},
  issue = {21},
  pages = {214407},
  numpages = {8},
  year = {2018},
  month = {Dec},
  publisher = {American Physical Society},
  doi = {10.1103/PhysRevB.98.214407},
  url = {https://link.aps.org/doi/10.1103/PhysRevB.98.214407}
}

@article{Denisov-jpcc,
doi = {10.1088/1361-648X/ab966e},
url = {https://doi.org/10.1088/1361-648X/ab966e},
year = {2020},
month = {jul},
publisher = {IOP Publishing},
volume = {32},
number = {41},
pages = {415302},
author = {Denisov, K S},
title = {Theory of an electron asymmetric scattering on skyrmion textures in two-dimensional systems},
journal = {Journal of Physics: Condensed Matter}
}

@article{Rashba-Ramsauer,
  title = {Giant resonances in topological spin {Hall} effect due to electron-skyrmion scattering in two-dimensional {Rashba} spin-orbit ferromagnets},
  author = {Zadorozhnyi, Andrei and Dahnovsky, Yuri},
  journal = {Phys. Rev. B},
  volume = {105},
  issue = {1},
  pages = {014445},
  numpages = {11},
  year = {2022},
  month = {Jan},
  publisher = {American Physical Society},
  doi = {10.1103/PhysRevB.105.014445},
  url = {https://link.aps.org/doi/10.1103/PhysRevB.105.014445}
}

@article{Blugel-prl,
  title = {Transverse Transport in Two-Dimensional Relativistic Systems with Nontrivial Spin Textures},
  author = {Bouaziz, Juba and Ishida, Hiroshi and Lounis, Samir and Bl\"ugel, Stefan},
  journal = {Phys. Rev. Lett.},
  volume = {126},
  issue = {14},
  pages = {147203},
  numpages = {7},
  year = {2021},
  month = {Apr},
  publisher = {American Physical Society},
  doi = {10.1103/PhysRevLett.126.147203},
  url = {https://link.aps.org/doi/10.1103/PhysRevLett.126.147203}
}

@article{Hareram,
  title = {Quantum dynamics of electron scattering from skyrmions},
  author = {Swain, Hareram and Mandal, Arijit and Satpathy, S. and Nanda, B. R. K.},
  journal = {Phys. Rev. B},
  volume = {113},
  issue = {10},
  pages = {104446},
  numpages = {13},
  year = {2026},
  month = {Mar},
  publisher = {American Physical Society},
  doi = {10.1103/2pyv-dcl9},
  url = {https://link.aps.org/doi/10.1103/2pyv-dcl9}
}

@article{Mandal,
  title = {Tuning the band topology and topological {Hall} effect in skyrmion crystals via the spin-orbit coupling},
  author = {Mandal, Arijit and Satpathy, S. and Nanda, B. R. K.},
  journal = {Phys. Rev. B},
  volume = {112},
  issue = {6},
  pages = {064428},
  numpages = {13},
  year = {2025},
  month = {Aug},
  publisher = {American Physical Society},
  doi = {10.1103/tlpn-9s4z},
  url = {https://link.aps.org/doi/10.1103/tlpn-9s4z}
}

@article{Nagaosa_the,
  title = {Quantized topological {Hall} effect in skyrmion crystal},
  author = {Hamamoto, Keita and Ezawa, Motohiko and Nagaosa, Naoto},
  journal = {Phys. Rev. B},
  volume = {92},
  issue = {11},
  pages = {115417},
  numpages = {6},
  year = {2015},
  month = {Sep},
  publisher = {American Physical Society},
  doi = {10.1103/PhysRevB.92.115417},
  url = {https://link.aps.org/doi/10.1103/PhysRevB.92.115417}
}

@article{Goebel_spin,
  title = {Antiferromagnetic skyrmion crystals: Generation, topological {Hall}, and topological spin {Hall} effect},
  author = {G\"obel, B\"orge and Mook, Alexander and Henk, J\"urgen and Mertig, Ingrid},
  journal = {Phys. Rev. B},
  volume = {96},
  issue = {6},
  pages = {060406(R)},
  numpages = {5},
  year = {2017},
  month = {Aug},
  publisher = {American Physical Society},
  doi = {10.1103/PhysRevB.96.060406},
  url = {https://link.aps.org/doi/10.1103/PhysRevB.96.060406}
}

@Article{Goebel_orbital,
author={G{\"o}bel, B{\"o}rge
and Schimpf, Lennart
and Mertig, Ingrid},
title={Topological orbital {Hall} effect caused by skyrmions and antiferromagnetic skyrmions},
journal={Communications Physics},
year={2025},
month={Jan},
day={11},
volume={8},
number={1},
pages={17},
doi={10.1038/s42005-024-01925-x},
url={https://doi.org/10.1038/s42005-024-01925-x}
}

@article{Lapidus,
    author = {Lapidus, I. Richard},
    title = {Quantum‐mechanical scattering in two dimensions},
    journal = {American Journal of Physics},
    volume = {50},
    number = {1},
    pages = {45-47},
    year = {1982},
    month = {01},
    issn = {0002-9505},
    doi = {10.1119/1.13004},
    url = {https://doi.org/10.1119/1.13004}
}

@misc{Note2,
      note = {It is easy to see that the exchange interaction $J$ does not appear in the unperturbed GF, Eq. (6).  This is because both the energy of the incoming particle $E = \hbar^2 k_\alpha^2 / 2m + J \ \delta_{\alpha \downarrow}$ and the unperturbed Hamiltonian ${\mathcal {H}}_0$ in Eq. (2)  include $J \ \delta_{\alpha \downarrow}$, which therefore cancels out in the GF expression $G_0 (E) = (E-{\mathcal {H}}_0+i \eta)^{-1} $. This leaves  $k_\alpha$ as the sole parameter to differentiate between the two spin channels in Eq. (6).}
    }

@article{thc-review,
title = {Topological {H}all transport: Materials, mechanisms and potential applications},
journal = {Progress in Materials Science},
volume = {130},
pages = {100971},
year = {2022},
issn = {0079-6425},
doi = {https://doi.org/10.1016/j.pmatsci.2022.100971},
url = {https://www.sciencedirect.com/science/article/pii/S0079642522000524},
author = {Han Wang and Yingying Dai and Gan-Moog Chow and Jingsheng Chen}
}

@article{nagaosa-review,
title={Topological properties and dynamics of magnetic skyrmions},
author={Nagaosa, Naoto and Tokura, Yoshinori},
journal={Nature Nanotechnology},
year={2013},
month={Dec},
day={01},
volume={8},
number={12},
pages={899-911},
issn={1748-3395},
doi={10.1038/nnano.2013.243},
url={https://doi.org/10.1038/nnano.2013.243}
}

@article{tokura-review,
title={Magnetic Skyrmion Materials},
author={Tokura, Yoshinori and Kanazawa, Naoya},
journal={Chemical Reviews},
year={2021},
month={Mar},
day={10},
publisher={American Chemical Society},
volume={121},
number={5},
pages={2857-2897},
issn={0009-2665},
doi={10.1021/acs.chemrev.0c00297},
url={https://doi.org/10.1021/acs.chemrev.0c00297}
}

@article{Beyond,
title = {Beyond skyrmions: Review and perspectives of alternative magnetic quasiparticles},
journal = {Physics Reports},
volume = {895},
pages = {1-28},
year = {2021},
issn = {0370-1573},
doi = {https://doi.org/10.1016/j.physrep.2020.10.001},
url = {https://www.sciencedirect.com/science/article/pii/S0370157320303525},
author = {Börge Göbel and Ingrid Mertig and Oleg A. Tretiakov}
}

@article{the1,
  title = {Topological {Hall} Effect in the {$A$} Phase of {MnSi}},
  author = {Neubauer, A. and Pfleiderer, C. and Binz, B. and Rosch, A. and Ritz, R. and Niklowitz, P. G. and B\"oni, P.},
  journal = {Phys. Rev. Lett.},
  volume = {102},
  issue = {18},
  pages = {186602},
  numpages = {4},
  year = {2009},
  month = {May},
  publisher = {American Physical Society},
  doi = {10.1103/PhysRevLett.102.186602},
  url = {https://link.aps.org/doi/10.1103/PhysRevLett.102.186602}
}

@article{the2,
  title = {Large Topological {Hall} Effect in a Short-Period Helimagnet {MnGe}},
  author = {Kanazawa, N. and Onose, Y. and Arima, T. and Okuyama, D. and Ohoyama, K. and Wakimoto, S. and Kakurai, K. and Ishiwata, S. and Tokura, Y.},
  journal = {Phys. Rev. Lett.},
  volume = {106},
  issue = {15},
  pages = {156603},
  numpages = {4},
  year = {2011},
  month = {Apr},
  publisher = {American Physical Society},
  doi = {10.1103/PhysRevLett.106.156603},
  url = {https://link.aps.org/doi/10.1103/PhysRevLett.106.156603}
}

@article{the3,
  title = {Robust Formation of Skyrmions and Topological {Hall} Effect Anomaly in Epitaxial Thin Films of {MnSi}},
  author = {Li, Yufan and Kanazawa, N. and Yu, X. Z. and Tsukazaki, A. and Kawasaki, M. and Ichikawa, M. and Jin, X. F. and Kagawa, F. and Tokura, Y.},
  journal = {Phys. Rev. Lett.},
  volume = {110},
  issue = {11},
  pages = {117202},
  numpages = {5},
  year = {2013},
  month = {Mar},
  publisher = {American Physical Society},
  doi = {10.1103/PhysRevLett.110.117202},
  url = {https://link.aps.org/doi/10.1103/PhysRevLett.110.117202}
}

@article{the4,
  title = {Unusual {Hall} Effect Anomaly in {MnSi} under Pressure},
  author = {Lee, Minhyea and Kang, W. and Onose, Y. and Tokura, Y. and Ong, N. P.},
  journal = {Phys. Rev. Lett.},
  volume = {102},
  issue = {18},
  pages = {186601},
  numpages = {4},
  year = {2009},
  month = {May},
  publisher = {American Physical Society},
  doi = {10.1103/PhysRevLett.102.186601},
  url = {https://link.aps.org/doi/10.1103/PhysRevLett.102.186601}
}

@article{RT-expt,
  title = {Low-Energy ${e}^{\ensuremath{-}}$-{A}r Total Scattering Cross Sections: The {Ramsauer-Townsend} Effect},
  author = {Golden, D. E. and Bandel, H. W.},
  journal = {Phys. Rev.},
  volume = {149},
  issue = {1},
  pages = {58--59},
  numpages = {0},
  year = {1966},
  month = {Sep},
  publisher = {American Physical Society},
  doi = {10.1103/PhysRev.149.58},
  url = {https://link.aps.org/doi/10.1103/PhysRev.149.58}
}

@article{Ramsaur_1921,
  title={{\"U}ber den Wirkungsquerschnitt der Gasmolek{\"u}le gegen{\"u}ber langsamen Elektronen},
  author={Ramsauer, Carl},
  journal={Annalen der Physik},
  volume={369},
  number={6},
  pages={513--540},
  year={1921},
  publisher={WILEY-VCH Verlag Leipzig}
}

@article{Townsend_1922,
  title={The motion of electrons in argon},
  author={Townsend, JS and Bailey, VA},
  journal={Phil. Mag.},
  volume={43},
  number={255},
  pages={593--600},
  year={1922},
  publisher={Taylor \& Francis}
}

\end{document}